\documentclass[12pt,twoside, a4paper]{article}
\def\pd{\partial}
\def\mc{\mathcal}
\def\ul{\underline}

\usepackage[dvips]{graphicx}
\usepackage{amssymb}
\usepackage[bbgreekl]{mathbbol}
\usepackage{amssymb,amsmath}
\usepackage{graphicx}
\usepackage{dsfont}
\usepackage{caption}
\usepackage{subcaption}
\usepackage{mathtools}
\usepackage{verbatim}
\usepackage{graphicx}
\usepackage{multirow}
\usepackage[outline]{contour}
\usepackage{xcolor,colortbl}
\input{epsf.sty}  
\begin{document}
\begin{center}
\Large{\textbf{Line and surface defects in 5D $N=2$ SCFT from matter-coupled $F(4)$ gauged supergravity}}
\end{center}
\vspace{1 cm}
\begin{center}
\large{\textbf{Parinya Karndumri}}
\end{center}
\begin{center}
String Theory and Supergravity Group, Department
of Physics, Faculty of Science, Chulalongkorn University, 254 Phayathai Road, Pathumwan, Bangkok 10330, Thailand \\
E-mail: parinya.ka@hotmail.com \vspace{1 cm}
\end{center}
\begin{abstract}
We study supersymmetric solutions of matter-coupled $F(4)$ gauged supergravity in the forms of $AdS_2\times S^3$- and $AdS_3\times S^2$-sliced domain walls with a non-vanishing two-form field from the supergravity multiplet. These two types of solutions holographically describe conformal line and surface defects within five-dimensional $N=2$ SCFTs, respectively. In the case of $SO(3)\times SO(3)$ gauge group obtained by coupling $F(4)$ supergravity to three vector multiplets, we find charged domain wall solutions describing holographic RG flows between two supersymmetric $AdS_6$ vacua with a running two-form field supporting the line and surface defects. For a particular case of one vector multiplet with $SO(3)\times U(1)$ gauge group, we find solutions describing RG flows from an $AdS_6$ vacuum to physically acceptable singular geometries in the presence of defects. This class of solutions can be uplifted to type IIB theory using a consistent truncation obtained from $SO(5,5)$ exceptional field theory with half-maximal structures.   
\end{abstract}
\newpage
%%%%%%%%%%%%%%%%%%%%%%%%%%%%%%%%%%%%%%%%%%%%%%%%%%%%%%%%%%%%%%%%%%%%%%%%%%%%%%%%%%%%%%%%%%%%%%%%%%%%%%%%%%%%%%%%%%%%%%%%%%%%%%%%%%%%%%%%%
\section{Introduction}
One of many interesting results that can be holographically studied via the AdS/CFT correspondence \cite{maldacena,Gubser_AdS_CFT,Witten_AdS_CFT} is the study of conformal defects within higher-dimensional superconformal field theories (SCFTs), see for example \cite{defect1,defect2,defect3,boundary_defect,N08_defect,quantum_AdS_dCFT1,quantum_AdS_dCFT2,surface_AdS3,AdS2_4D_defect,AdS2_defect,correlator_defect}. In particular, for SCFTs in five and six dimensions for which there is no Lagrangian description available, various insights into strongly-coupled dynamics have come mainly from the holographic study of the proposed gravity duals. Up to now, most of the interesting results have arisen fruitfully from studying solutions of lower-dimensional gauged supergravity. A number of complications present in the higher-dimensional computations become more traceable in the framework of gauged supergravity. Furthermore, a number of holographic computations are also simpler in gauged supergravity setting. In some cases, the solutions can be uplifted to ten or eleven dimensions via consistent truncations generating new holographic solutions in the context of string/M-theory. In other cases for which embedding in string/M-theory is presently not known, various types of holographic solutions within gauged supergravity are still worth studying.  
\\
\indent In this paper, we are interested in conformal line and surface defects within five-dimensional $N=2$ SCFTs in the framework of matter-coupled $F(4)$ gauged supergravity \cite{F4SUGRA1,F4SUGRA2}. These defects preserve one- and two-dimensional conformal symmetry subgroup of the $SO(2,5)$ conformal symmetry in five dimensions. The corresponding supergravity solutions can be constructed by taking the metric ansatz in the form of curved domain walls with $AdS_2\times S^3$ and $AdS_3\times S^2$ slicing, respectively. These types of solutions have been considered in the context of pure $F(4)$ gauged supergravity, constructed in \cite{F4_Romans}, in \cite{AdS2_F4_Dibitetto} and \cite{surface_defect_F4_Dibitetto}, see also a more general solution in \cite{line_defect_F4_Gutperle}. In this case, the solutions can be uplifted to massive type IIA theory using the consistent truncation constructed in \cite{F4_form_mIIA}, for other holographic studies of conformal defects in five-dimensional SCFTs see \cite{Christoph1,Christoph2,Christoph3}.
\\
\indent In this work, we generalize these two classes of solutions to the matter-coupled $F(4)$ gauged supergravity. We will consider two gauged supergravities obtained from coupling pure or minimal $F(4)$ gauged supergravity to three and one vector multiplets. The former leads to $SO(4)\sim SO(3)_R\times SO(3)$ gauge group admitting two supersymmetric $AdS_6$ vacua with $SO(3)_R\times SO(3)$ and $SO(3)_{\textrm{diag}}$ symmetries \cite{F4_flow}, see also \cite{5DSYM_from_F4,6D_twist,AdS6_BH,6D_Janus_RG} for other hologrpahic solutions obtained from this gauged supergravity. $SO(3)_R$ denotes the R-symmetry while the second $SO(3)$ factor arises from the symmetry of the three vector multiplets. On the other hand, the latter gauged supergravity leads to $SO(3)_R\times U(1)$ gauge group. The $U(1)$ factor is just the abelian gauge symmetry of the single vector multiplet under which all the other fields are uncharged. A number of supersymmetric Janus solutions and $AdS_6$ black holes from this gauged supergravity has been studied more recently in \cite{6D_Janus,AdS6_BH_Zaffaroni,AdS6_BH_Minwoo}. For similar holographic solutions describing conformal surface defects in seven-dimensional gauged supergravities, see \cite{7D_sol_Dibitetto,7D_N2_DW_3_form,6D_surface_Dibitetto,7D_N4_DW_3_form,7D_defect_Nicolo}.
\\
\indent In the case of $SO(3)_R\times SO(3)$ $F(4)$ gauged supergravity, we consider line and surface defect solutions in the presence of two five-dimensional conformal fixed points dual to the two supersymmetric $AdS_6$ vacua. This type of solutions has not previously appeared since the pure $F(4)$ gauged supergravity admits only one supersymmetric $AdS_6$ vacuum. The solutions interpolate between the two aforementioned $AdS_6$ vacua with a running two-form field, generalizing the holographic RG flow studied in \cite{F4_flow} by including the two-form field. For $SO(3)_R\times U(1)$ gauged supergravity, it has been shown in \cite{Henning_Malek_AdS7_6} that this gauged supergravity can be obtained from a consistent truncation of type IIB theory on a product of a two-sphere and a Riemann surface $(S^2\times \Sigma)$. In this case, the solutions are similar to those studied in \cite{AdS2_F4_Dibitetto,surface_defect_F4_Dibitetto} and interpolate between an $AdS_6$ vacuum and singular geometries due to the non-existence of the second supersymmetric $AdS_6$ critical point of the scalar potential. We will also uplift the resulting solutions to type IIB theory using the truncation formulae given in \cite{Henning_Malek_AdS7_6}.
\\
\indent The paper is organized as follows. We give a very brief review of matter-coupled $F(4)$ gauged supergravity as constructed in \cite{F4SUGRA1, F4SUGRA2} in section \ref{6D_SO4gaugedN2}. A number of new solutions describing line and surface defects in the forms of $AdS_2\times S^3$- and $AdS_3\times S^2$-sliced domain walls are given respectively in sections \ref{line_defects} and \ref{surface_defects}. We also give uplifted solutions to type IIB theory for the case of one vector multiplet and $SO(3)\times U(1)$ gauge group. We give some conclusions and comments in section \ref{conclusion}. In the appendices, we give some details on the derivation of the BPS equations and collect relevant formulae in the consistent truncation of type IIB theory on $S^2\times \Sigma$. 
%%%%%%%%%%%%%%%%%%%%%%%%%%%%%%%%%%%%%%%%%%%%%%%%%%%%%%%%%%%%%%%%%%%%%%%%%%%%%%%%%%%%%%%%%%%%%%%%%%%%%%%%%%%%%%%%%%%%%%%%%%%%%%%%%%%%%%%%%
\section{Matter coupled $N=(1,1)$ gauged supergravity in six dimensions}\label{6D_SO4gaugedN2}
We first review the structure of matter-coupled $F(4)$ gauged supergravity in six dimensions coupled to $n$ vector multiplets. The supergravity and $n$ vector multiplets have the component fields 
\begin{equation}
\left(e^{\hat{{\mu}}}_\mu,\psi^A_\mu, A^\alpha_\mu, B_{\mu\nu}, \chi^A,
\sigma\right)\qquad \textrm{and}\qquad (A_\mu,\lambda_A,\phi^\alpha)^I\, .
\end{equation}
We use the metric signature $(-+++++)$ with space-time and tangent space (flat) indices denoted respectively by $\mu,\nu=0,\ldots ,5$ and $\hat{\mu},\hat{\nu}=0,\ldots, 5$. The bosonic fields are the graviton $e^{\hat{\mu}}_\mu$, a two-form field $B_{\mu\nu}$, $4+n$ vector fields $A^\Lambda=(A^\alpha_\mu,A^I)$, with $\alpha=0,1,2,3$ and $I=1,2,\ldots, n$, together with the dilaton $\sigma$ and $4n$ scalars $\phi^{\alpha I}$. The fermionic fields are given by two gravitini $\psi^A_\mu$, two spin-$\frac{1}{2}$ fields $\chi^A$ and $2n$ gaugini $\lambda^I_A$ with indices $A,B,\ldots =1,2$ denoting the fundamental representation of $SU(2)_R\sim USp(2)_R\sim SO(3)_R$ R-symmetry. We will also split $\alpha, \beta$ indices as $\alpha=(0,r)$ with $r,s,\ldots =1,2,3$ being $SU(2)_R$ adjoint indices.
\\
\indent The dilaton and $4n$ scalars from the vector multiplets are described by $\mathbb{R}^+\times SO(4,n)/SO(4)\times SO(n)$ coset with $\mathbb{R}^+$ corresponding to the dilaton. The second factor can be parametrized by a coset
representative ${L^\Lambda}_{\ul{\Sigma}}$ transforming under the global $SO(4,n)$ and local $SO(4)\times SO(n)$ symmetries by left and right multiplications respectively with indices $\Lambda,\ul{\Sigma}=0,\ldots , n+3$. We can also split the local index as $\ul{\Sigma}=(\alpha,I)=(0,r,I)$ leading to various components of the coset representative of the form
\begin{equation}
{L^\Lambda}_{\ul{\Sigma}}=({L^\Lambda}_\alpha,{L^\Lambda}_I).
\end{equation}
The inverse of ${L^\Lambda}_{\ul{\Sigma}}$ will be denoted by ${(L^{-1})^{\ul{\Lambda}}}_\Sigma=({(L^{-1})^{\alpha}}_\Sigma,{(L^{-1})^{I}}_\Sigma)$. $SO(4,n)$ indices are raised and lowered by the invariant tensor
\begin{equation}
\eta_{\Lambda\Sigma}=\eta^{\Lambda\Sigma}=(\delta_{\alpha\beta},-\delta_{IJ}).
\end{equation}
\indent The bosonic Lagrangian of the matter-coupled gauged supergravity can be written as
\begin{eqnarray}
e^{-1}\mathcal{L}&=&\frac{1}{4}R-\pd_\mu \sigma\pd^\mu \sigma
-\frac{1}{4}P^{I\alpha}_\mu P^{\mu}_{I\alpha}-\frac{1}{8}e^{-2\sigma}\mc{N}_{\Lambda\Sigma}\widehat{F}^\Lambda_{\mu\nu}
\widehat{F}^{\Sigma\mu\nu}-\frac{3}{64}e^{4\sigma}H_{\mu\nu\rho}H^{\mu\nu\rho}\nonumber \\
& &-V-\frac{1}{64}e^{-1}\epsilon^{\mu\nu\rho\sigma\lambda\tau}B_{\mu\nu}\left(\eta_{\Lambda\Sigma}\widehat{F}^\Lambda_{\rho\sigma}
\widehat{F}^\Sigma_{\lambda\tau}+mB_{\rho\sigma}\widehat{F}^\Lambda_{\lambda\tau}\delta^{\Lambda 0}+\frac{1}{3}m^2B_{\rho\sigma}B_{\lambda\tau}\right)\nonumber \\
\label{Lar}
\end{eqnarray}
with $e=\sqrt{-g}$. The corresponding field strength tensors are defined by
\begin{eqnarray}
\widehat{F}^\Lambda=F^\Lambda-m\delta^{\Lambda 0}B, \qquad F^\Lambda=dA^\Lambda+\frac{1}{2}{f^\Lambda}_{\Sigma\Gamma} A^\Sigma\wedge A^\Gamma,\qquad H=dB\, .
\end{eqnarray}
${f^\Lambda}_{\Sigma\Gamma}$ are structure constants of the gauge group. We also note the convention on differential forms used in \cite{F4SUGRA1, F4SUGRA2}
\begin{equation}
F^\Lambda=F^\Lambda_{\mu\nu} dx^\mu \wedge dx^\nu\qquad \textrm{and}\qquad H=H_{\mu\nu\rho}dx^\mu \wedge dx^\nu\wedge dx^\rho
\end{equation}
which are different from the usual convention. 
\\
\indent The vielbein on $SO(4,n)/SO(4)\times SO(n)$ denoted by $P^{I\alpha}_\mu=P^{I\alpha}_x\pd_\mu\phi^x$, $x=1,\ldots, 4n$, can be obtained from the left-invariant 1-form
\begin{equation}
{\Omega^{\ul{\Lambda}}}_{\ul{\Sigma}}=
{(L^{-1})^{\ul{\Lambda}}}_{\Pi}\nabla {L^\Pi}_{\ul{\Sigma}}
\qquad \textrm{with}\qquad \nabla
{L^\Lambda}_{\ul{\Sigma}}={dL^\Lambda}_{\ul{\Sigma}}
-f^{\phantom{\Gamma}\Lambda}_{\Gamma\phantom{\Lambda}\Pi}A^\Gamma
{L^\Pi}_{\ul{\Sigma}}\, .
\end{equation}
as  
\begin{equation}
{P^I}_{\alpha}=({P^I}_{0},{P^I}_{r})=(\Omega^I_{\phantom{a}0},\Omega^I_{\phantom{a}r}).
\end{equation}
The remaining components of ${\Omega^{\ul{\Lambda}}}_{\ul{\Sigma}}$ are identified as the $SO(4)\times SO(n)$ composite connections $(\Omega^{rs},\Omega^{r0},\Omega^{IJ})$. The symmetric scalar matrix $\mc{N}_{\Lambda\Sigma}$ appearing in the vector kinetic term is defined by
\begin{eqnarray}
\mc{N}_{\Lambda\Sigma}=L_{\Lambda\alpha}{(L^{-1})^\alpha}_\Sigma-L_{\Lambda I}{(L^{-1})^I}_\Sigma=(\eta L L^T\eta)_{\Lambda\Sigma}\, .
\end{eqnarray}
The scalar potential is given by
\begin{eqnarray}
V&=&-e^{2\sigma}\left[\frac{1}{36}A^2+\frac{1}{4}B^iB_i+\frac{1}{4}\left(C^I_{\phantom{s}t}C_{It}+4D^I_{\phantom{s}t}D_{It}\right)\right]
+m^2e^{-6\sigma}\mc{N}_{00}\nonumber \\
& &-me^{-2\sigma}\left[\frac{2}{3}AL_{00}-2B^iL_{0i}\right].
\end{eqnarray}
We are only interested in compact gauge goups of the form $SU(2)_R\times G_c$ with $G_c\subset SO(n)$ and ${f^\Lambda}_{\Sigma\Gamma}=(g_1\epsilon_{rst},g_2C_{IJK})$. $C_{IJK}$ are structure constants of $G_c$. Various components of fermion-shift matrices are defined by 
\begin{eqnarray}
A&=&\epsilon^{rst}K_{rst},\qquad B^i=\epsilon^{ijk}K_{jk0},\\
C^{\phantom{ts}t}_I&=&\epsilon^{trs}K_{rIs},\qquad D_{It}=K_{0It}
\end{eqnarray}
in which
\begin{eqnarray}
K_{rst}&=&g_1\epsilon_{lmn}L^l_{\phantom{r}r}(L^{-1})_s^{\phantom{s}m}L_{\phantom{s}t}^n+
g_2C_{IJK}L^I_{\phantom{r}r}(L^{-1})_s^{\phantom{s}J}L_{\phantom{s}t}^K,\nonumber
\\
K_{rs0}&=&g_1\epsilon_{lmn}L^l_{\phantom{r}r}(L^{-1})_s^{\phantom{s}m}L_{\phantom{s}0}^n+
g_2C_{IJK}L^I_{\phantom{r}r}(L^{-1})_s^{\phantom{s}J}L_{\phantom{s}0}^K,\nonumber
\\
K_{rIt}&=&g_1\epsilon_{lmn}L^l_{\phantom{r}r}(L^{-1})_I^{\phantom{s}m}L_{\phantom{s}t}^n+
g_2C_{IJK}L^I_{\phantom{r}r}(L^{-1})_I^{\phantom{s}J}L_{\phantom{s}t}^K,\nonumber
\\
K_{0It}&=&g_1\epsilon_{lmn}L^l_{\phantom{r}0}(L^{-1})_I^{\phantom{s}m}L_{\phantom{s}t}^n+
g_2C_{IJK}L^I_{\phantom{r}0}(L^{-1})_I^{\phantom{s}J}L_{\phantom{s}t}^K\,
.
\end{eqnarray}
Supersymmetry transformations of the fermionic fields are given by
\begin{eqnarray}
\delta\psi_{\mu
A}&=&D_\mu\epsilon_A-\frac{1}{24}\left(Ae^\sigma+6me^{-3\sigma}(L^{-1})_{00}\right)\epsilon_{AB}\gamma_\mu\epsilon^B\nonumber
\\
& &-\frac{1}{8}
\left(B_te^\sigma-2me^{-3\sigma}(L^{-1})_{t0}\right)\gamma^7\sigma^t_{AB}\gamma_\mu\epsilon^B\nonumber \\
&
&+\frac{i}{16}e^{-\sigma}\left[\epsilon_{AB}(L^{-1})_{0\Lambda}\gamma_7+\sigma^r_{AB}(L^{-1})_{r\Lambda}\right]
F^\Lambda_{\nu\lambda}(\gamma_\mu^{\phantom{s}\nu\lambda}
-6\delta^\nu_\mu\gamma^\lambda)\epsilon^B\nonumber \\
& &+\frac{i}{32}e^{2\sigma}H_{\nu\lambda\rho}\gamma_7({\gamma_\mu}^{\nu\lambda\rho}-3\delta_\mu^\nu\gamma^{\lambda\rho})\epsilon_A,\label{delta_psi}\\
\delta\chi_A&=&\frac{1}{2}\gamma^\mu\pd_\mu\sigma\epsilon_{AB}\epsilon^B+\frac{1}{24}
\left[Ae^\sigma-18me^{-3\sigma}(L^{-1})_{00}\right]\epsilon_{AB}\epsilon^B\nonumber
\\
& &-\frac{1}{8}
\left[B_te^\sigma+6me^{-3\sigma}(L^{-1})_{t0}\right]\gamma^7\sigma^t_{AB}\epsilon^B\nonumber
\\
& &-\frac{i}{16}e^{-\sigma}\left[\sigma^r_{AB}(L^{-1})_{r\Lambda}-\epsilon_{AB}(L^{-1})_{0\Lambda}\gamma_7\right]F^\Lambda_{\mu\nu}\gamma^{\mu\nu}\epsilon^B\nonumber \\
& &-\frac{i}{32}e^{2\sigma}H_{\nu\lambda\rho}\gamma_7\gamma^{\nu\lambda\rho}\epsilon_A,
\label{delta_chi}\\
%\end{eqnarray}
%\begin{eqnarray}
\delta
\lambda^{I}_A&=&P^I_{ri}\gamma^\mu\pd_\mu\phi^i\sigma^{r}_{\phantom{s}AB}\epsilon^B+P^I_{0i}
\gamma^7\gamma^\mu\pd_\mu\phi^i\epsilon_{AB}\epsilon^B-\left(2i\gamma^7D^I_{\phantom{s}t}+C^I_{\phantom{s}t}\right)
e^\sigma\sigma^t_{AB}\epsilon^B \nonumber
\\
& &+2me^{-3\sigma}(L^{-1})^I_{\phantom{ss}0}
\gamma^7\epsilon_{AB}\epsilon^B-\frac{i}{2}e^{-\sigma}(L^{-1})^I_{\phantom{s}\Lambda}F^\Lambda_{\mu\nu}
\gamma^{\mu\nu}\epsilon_{A}\label{delta_lambda}
\end{eqnarray}
with ${\sigma^{rA}}_B$ being Pauli matrices. $SU(2)_R$ fundamental indices $A,B$ can be raised and lowered by $\epsilon^{AB}$ and $\epsilon_{AB}$ with the convention $T^A=\epsilon^{AB}T_B$ and $T_A=T^B\epsilon_{BA}$. Finally, the covariant derivative of $\epsilon_A$ is given by
\begin{equation}
D_\mu \epsilon_A=\pd_\mu
\epsilon_A+\frac{1}{4}\omega_\mu^{ab}\gamma_{ab}\epsilon_A+\frac{i}{2}\sigma^r_{AB}
\left[\frac{1}{2}\epsilon^{rst}\Omega_{\mu st}-i\gamma_7
\Omega_{\mu r0}\right]\epsilon^B\, .
\end{equation}

%%%%%%%%%%%%%%%%%%%%%%%%%%%%%%%%%%%%%%%%%%%%%%%%%%%%%%%%%%%%%%%%%%%%%%%%%%%%%%%%%%%%%%%%%%%%%%%%%%%%%%%%%%%%%%%%%%%%%%%%%%%%%%%%%%%%%%%%%
\section{Line defects from matter-coupled $F(4)$ gauged supergravity}\label{line_defects}
In this section, we consider solutions of the matter-coupled $F(4)$ gauged supergravity which are holographically dual to conformal line defects within $N=2$ SCFTs in five dimensions. The procedure to find these solutions closely follows that given in \cite{AdS2_F4_Dibitetto}. The metric ansatz can be written as
\begin{equation}
ds^2=e^{2f(r)}ds^2_{AdS_2}+dr^2+e^{2h(r)}ds^2_{S^3}\, .\label{line_metric}
\end{equation}
This metric takes the form of a curved $AdS_2\times S^3$-sliced domain wall supported by a non-vanishing two-form field. We will also split the six-dimensional coordinates as $x^\mu =(x^\alpha,r,\theta^i)$ for $\alpha=0,1$ and $i=1,2,3$. Following \cite{AdS2_F4_Dibitetto}, we will take the ansatz for the two-form field to be 
\begin{equation}
B=b(r)\textrm{vol}_{AdS_2}\, .
\end{equation}
We will not consider solutions with non-vanishing vector fields in this paper. As pointed out in \cite{surface_defect_F4_Dibitetto}, in this case, the action of supersymmetry transformations on $SU(2)_R$ indices $A,B$ is trivial, so we can combine the two chiralities of the Symplectic-Majorana-Weyl spinors into a Dirac spinor. A suitable ansatz for the Killing spinors takes the form
\begin{equation}
\epsilon=\epsilon^++\epsilon^-
\end{equation}
with
\begin{eqnarray}
& &\epsilon^+=iY(r)\left[\cos\theta(r)\chi^+\otimes \epsilon_0+\sin\theta(r)\chi^+\otimes \sigma_3\epsilon_0\right]\otimes \eta,\nonumber \\
& &\epsilon^-=Y(r)\left[\sin\theta(r)\chi^-\otimes \epsilon_0-\cos\theta(r)\chi^-\otimes \sigma_3\epsilon_0\right]\otimes \eta\, .
\end{eqnarray}  
$\epsilon_0$ is a two-component constant spinor while $\eta$ and $\chi^\pm$ are respectively Killing spinors on $AdS_2$ and $S^3$ satisfying   
\begin{equation}
\hat{\nabla}_\alpha\chi^\pm=\pm\frac{i}{2L}\rho_\alpha\chi^\mp\qquad \textrm{and}\qquad \widetilde{\nabla}_i\eta=\frac{i}{2R}\tilde{\gamma}_i\eta\, .\label{AdS2_S3_Killing}
\end{equation}
$L$ and $R$ denote the radii of $AdS_2$ and $S^3$, respectively.  
\\
\indent Due to some differences in the convention for $F(4)$ gauged supergravity of \cite{F4_Romans} used in \cite{AdS2_F4_Dibitetto} and that of \cite{F4SUGRA1,F4SUGRA2} used here, we choose a slightly different choice of gamma matrices
\begin{eqnarray}
& &\gamma^\alpha=\rho^\alpha\otimes \sigma_2\otimes\mathbb{I}_2,\qquad \gamma^r=\rho_*\otimes \sigma_2\otimes \mathbb{I}_2,\nonumber \\
& &\gamma^i=-\mathbb{I}_2\otimes \sigma_1\otimes\tilde{\gamma}^i,\qquad \gamma_7=-i\mathbb{I}_2\otimes \sigma_3\otimes \mathbb{I}_2
\end{eqnarray}
which are related to those used in \cite{AdS2_F4_Dibitetto} by the relation $\gamma^\mu=-i\Gamma_*\Gamma^\mu$ with $\Gamma_*=i\gamma_7$. In the above equations, the chirality matrix on $AdS_2$ is defined by $\rho_*=\rho^0\rho^1$ with $\rho^2_*=\mathbb{I}_2$.
\\
\indent We will consider this type of solutions in matter-coupled $F(4)$ gauged supergravity for two different gauge groups. The first gauge group is given by $SO(3)_R\times SO(3)$ with the gauge structure constants
 \begin{equation}
{f^\Lambda}_{\Pi\Sigma}=(g_1\epsilon_{rst},g_2\epsilon_{IJK})
\end{equation}
with $g_1$ and $g_2$ being the corresponding gauge coupling constants for the two $SO(3)$ factors. This gauge group can be embedded in the case of $n\geq 3$ vector multiplets. For simplicity, we will consider only the simplest case of $n=3$ vector multiplets. The resulting gauged supergravity admits two supersymmetric $AdS_6$ vacua with $SO(3)_R\times SO(3)$ and $SO(3)_{\textrm{diag}}$ symmetries. Both of these vacua are maximally supersymmetric preserving the full sixteen supercharges.
\\
\indent The other gauge group under consideration here is $SO(3)_R\times U(1)\sim SU(2)_R\times U(1)$ obtained by coupling pure $F(4)$ gauged supergravity of \cite{F4_Romans} to one vector multiplet. The corresponding gauge structure constants are given by  
\begin{equation}
{f^\Lambda}_{\Pi\Sigma}=g_1\epsilon_{rst}\, . \label{SO2R_gauging}
\end{equation} 
It should be noted that the $U(1)$ factor is the usual abelian gauge symmetry of the additional vector field from the single vector multiplet as in the ungauged supergravity. Accordingly, all the other fields are uncharged under this $U(1)$ subgroup. It has been shown in \cite{Henning_Malek_AdS7_6} that this matter-coupled $F(4)$ gauged supergravity can be embedded in type IIB theory via a consistent truncation obtained from $SO(5,5)$ exceptional field theory with half-maximal structures.    

\subsection{Line defects with $SO(3)_{\textrm{diag}}$ symmetry}
We begin with the case of $F(4)$ gauged supergravity coupled to three vector multiplets with gauge group $SO(3)_R\times SO(3)$. We are interested in solutions with $SO(3)_{\textrm{diag}}$ symmetry. There is one singlet scalar from $SO(4,3)/SO(4)\times SO(3)$ coset \cite{F4_flow}. With the non-compact generators of $SO(4,3)$ written as
\begin{eqnarray}
Y_{\alpha I}=e^{\alpha,I+3}+e^{I+3,\alpha}
\end{eqnarray}
with   
\begin{equation}
(e^{\Lambda \Sigma})_{\Gamma \Pi}=\delta^\Lambda_{
\Gamma}\delta^\Sigma_{\Pi},\qquad \Lambda, \Sigma,\Gamma,
\Pi=0,\ldots ,6,
\end{equation}
the $SO(3)_{\textrm{diag}}$ invariant coset representative is given by
\begin{equation}
L=e^{\phi(Y_{11}+Y_{22}+Y_{33})}\, .\label{SO3diag_L}
\end{equation}
As identified in \cite{F4_flow}, there are two supersymmetric $AdS_6$ vacua with $SO(3)_R\times SO(3)$ and $SO(3)_{\textrm{diag}}$ symmetry in this sector. Since we are interested in solutions interpolating between these asymptotic $AdS_6$ geometries, it is useful to repeat these two vacua here
\begin{eqnarray}
\textrm{I}:& &\phi=0,\qquad \sigma=\frac{1}{4}\ln
\left[\frac{3m}{g_1}\right],\nonumber \\
& &V_0=-20m^2\left(\frac{g_1}{3m}\right)^{\frac{3}{2}},\qquad \ell_{\textrm{I}}=\frac{1}{2m}\left(\frac{3m}{g_1}\right)^{\frac{3}{4}},\label{SO4_AdS6}\\
\textrm{II}: & &\phi=\frac{1}{2}\ln \left[\frac{g_1+g_2}{g_2-g_1}\right],\qquad
\sigma=\frac{1}{4}\ln
\left[\frac{3m\sqrt{g_2^2-g_1^2}}{g_1g_2}\right],\nonumber \\
& &V_0=-20m^2\left[\frac{g_1g_2}{3m\sqrt{g_2^2-g_1^2}}\right]^{\frac{3}{2}},\qquad \ell_{\textrm{II}}=\frac{1}{2m}\left(\frac{3m\sqrt{g_2^2-g_1^2}}{g_1g_2}\right)^{\frac{3}{4}}\,
.\label{SO3_AdS6}
\end{eqnarray}
$V_0$ and $\ell$ are the cosmological constant and the $AdS_6$ radius with $\ell^2=-\frac{5}{V_0}$. We can also choose $g_1=3m$ to set $\sigma=0$ at the $SO(3)_R\times SO(3)$ critical point I.
\\
\indent We are now in a position to analyze relevant BPS equations. By imposing the projection condition on the Killing spinor of the form
\begin{equation}
\sigma_2\epsilon_0=\epsilon_0,
\end{equation} 
we find the following BPS equations
\begin{eqnarray}
f'&=&\frac{1}{4L}\left[e^{-f}\tan2\theta+Le^{-3\sigma}\sec2\theta\left[m-3m\cos4\theta\right.\right.\nonumber \\
& &\qquad  \left.\left.-2e^{4\sigma}(g_1\cosh^3\phi-g_2\sinh^3\phi)\right]\right], \\
h'&=&\frac{1}{4L}\left[e^{-f}\tan2\theta+Le^{-3\sigma}\sec2\theta\left[m\cos4\theta-3m\right.\right.\nonumber \\
& &\qquad \left.\left.-2e^{4\sigma}(g_1\cosh^3\phi-g_2\sinh^3\phi)\right]\right],\\
b'&=&-\frac{2}{L}e^{f-5\sigma}\sec2\theta\left[e^{3\sigma}+2e^fL\sin2\theta\left[2m\right.\right.\nonumber \\
& &\qquad \left.\left.+e^{4\sigma}(g_2\sinh^3\phi-g_1\cosh^3\phi)\right]\right],\qquad \\
\sigma'&=&-\frac{1}{4L}\sec2\theta\left[e^{-f}\sin2\theta+mLe^{-3\sigma}(5+\cos4\theta)\right.\nonumber \\
& &\qquad \left.-2Le^\sigma(g_1\cosh^3\phi-g_2\sinh^3\phi)\right],\\
Y'&=&-\frac{Y}{8L}\sec2\theta\left[e^{-f}\sin2\theta+mLe^{-3\sigma}(1-3\cos4\theta)\right.\nonumber \\
& &\qquad \left.-2Le^\sigma(g_1\cosh^3\phi-g_2\sinh^3\phi)\right],\\
\theta'&=&me^{-3\sigma}\sin2\theta,\\
0&=&\left[\phi'-e^\sigma\sinh2\phi(g_1\cosh\phi-g_2\sinh\phi)\right]\cos\theta,\label{eq_phi_1}\\
0&=&\left[\phi'+e^\sigma\sinh2\phi(g_1\cosh\phi-g_2\sinh\phi)\right]\sin\theta\label{eq_phi_2}
\end{eqnarray}
together with two algebraic constraints
\begin{eqnarray}
b&=&\frac{e^{f+\sigma}}{Lm}-2e^{2f-2\sigma}\sin2\theta,\\
R&=&\frac{1}{2}e^h\left[Le^{-f}\sec2\theta-2e^\sigma\tan2\theta(g_1\cosh^3\phi-g_2\sinh^3\phi)\right].
\end{eqnarray}
The detail on the derivation of these equations is given in appendix \ref{detail}. From equations \eqref{eq_phi_1} and \eqref{eq_phi_2}, we immediately see that we need to set $\sin\theta=0$ or $\cos\theta=0$ in order to have a non-trivial scalar profile from the vector multiplets. The two choices are equivalent and related by interchanging $r$ with $-r$. For definiteness, we will set $\cos\theta=0$ which leads to $f=h$. More precisely, either $\sin\theta=0$ or $\cos\theta=0$ give $h'=f'$ or $f=h+c$ for a constant $c$. However, we are interested in solutions with an asymptotically locally $AdS_6$ geometry with $f=h\sim \frac{r}{\ell}$ for $AdS_6$ radius $\ell$. We can then set $c=0$ and, with all these results, obtain a simpler set of BPS equations  
\begin{eqnarray} 
f'&=&\frac{1}{2}e^{-3\sigma}\left[m+e^{4\sigma}(g-1\cosh^3\phi-g_2\sinh^3\phi)\right],\\
\sigma'&=&\frac{3}{2}e^{-3\sigma}m+\frac{1}{2}e^\sigma\left(g_1\cosh^3\phi-g_2\sinh^3\phi\right),\\
\phi'&=&-e^\sigma\sin2\phi(g_1\cosh\phi-g_2\sinh\phi),\label{phi_eq}\\ 
b'&=&\frac{2e^{f-2\sigma}}{L},\qquad Y'=\frac{1}{2}Yf'\, .
\end{eqnarray}
The two algebraic constraints also simplify considerably 
\begin{eqnarray}
b=\frac{e^{f+\sigma}}{Lm}\qquad \textrm{and}\qquad L=2R\, .
\end{eqnarray}
It can be straightforwardly verified that the BPS equations are compatible with these two algebraic constraints and satisfy the second-order field equations.  
\\
\indent We note that the two-form field and the warp factor $f$ do not appear in other equations, so we can firstly solve for $\sigma$ and $\phi$ and eventually find $b$ and $f$. Solving for $\sigma$ in terms of $\phi$ as in \cite{F4_flow}, we find
\begin{equation}
\sigma=\frac{1}{4}\ln\left[\frac{\sigma_0+3m\coth\phi+3m\tanh\phi}{g_1\textrm{csch}\phi-g_2\textrm{sech}\phi}\right].
\end{equation}
As $\phi\rightarrow 0$, we see that $\sigma\sim \frac{1}{4}\ln\left[\frac{3m}{g_1}\right]$ which is the value of the dilaton at $AdS_6$ critical point I. We will determine the integration constant $\sigma_0$ by requiring that the solution interpolate between $AdS_6$ critical points I and II. This is achieved by setting
\begin{equation}
\sigma_0=-\frac{3m(g_1^2+g_2^2)}{g_1g_2}
\end{equation}
which gives
\begin{equation}
\sigma=\frac{1}{4}\ln\left[\frac{3m\cosh\phi}{g_1}-\frac{3m\sinh\phi}{g_2}\right].
\end{equation}
Similarly, we find
\begin{eqnarray}
f&=&\frac{1}{4}\ln(g_1\cosh\phi-g_2\sinh\phi)-\frac{1}{3}\ln\sinh2\phi \nonumber \\
& &+\frac{1}{12}\left[6m\cosh2\phi-\frac{3m(g_1^2+g_2^2)\sinh2\phi}{g_1g_2}\right],\\
b&=&\frac{3^{\frac{1}{3}}}{2^{\frac{1}{4}}L}\left[\frac{g_1^2+g_2^2-2g_1g_2\coth\phi-g_1g_2\tanh\phi}{g_1g_2m^2}\right]^{\frac{1}{3}},\\
Y&=&e^{\frac{f}{2}}\, .
\end{eqnarray}
Finally, we can use these results to find solution for $\phi$. By changing to a new radial coordinate $\tilde{r}$ given by $\frac{d\tilde{r}}{dr}=e^\sigma$, we can solve equation \eqref{phi_eq} to obtain
\begin{eqnarray}
2g_1g_2(\tilde{r}-\tilde{r}_0)&=&g_2\ln\cosh\frac{\phi}{2}-g_2\ln\sinh\frac{\phi}{2}-2g_1\tan^{-1}\tanh\frac{\phi}{2}\nonumber \\
& &-2\sqrt{g_2^2-g_1^2}\tanh^{-1}\left[\frac{g_2g_1\tanh\frac{\phi}{2}}{\sqrt{g_2^2-g_1^2}}\right].
\end{eqnarray}
\indent As $\phi\rightarrow 0$, for $g_1=3m$, we find $f\sim 2mr$ with $\tilde{r}\sim r$ and 
\begin{eqnarray}
\phi\sim e^{-6mr},\qquad \sigma\sim e^{-6mr},\qquad b\sim e^{2mr}
\end{eqnarray}
or in terms of the coordinate $z=e^{-\frac{r}{\ell_{\textrm{I}}}}$
\begin{eqnarray}
\phi\sim \sigma\sim z^3\qquad \textrm{and}\qquad b\sim z^{-1}\, .
\end{eqnarray}
Similarly, as $\phi\rightarrow \frac{1}{2}\ln\left[\frac{g_2+g_1}{g_2-g_1}\right]$, we find 
\begin{eqnarray}
\sigma\sim z^3,\qquad \phi\sim z^{-3},\qquad b\sim z^{-1}
\end{eqnarray}
with $z=e^{-\frac{r}{\ell_{\textrm{II}}}}$.
\\
\indent By linearizing the two-form field equation as in \cite{line_defect_F4_Gutperle}, we find that $b$ is dual to an operator of dimension $\Delta_b=4$ at both asymptotic $AdS_6$ geometries. Accordingly, the defect solution involves turning on a source of a dimension-$4$ operator dual to the two-form field. The solution is then a charged domain wall describing a holographic RG flow between two conformal fixed points of five-dimensional $N=2$ SCFT in the presence of a conformal line defect. It could be interesting to identify the field theory dual of this solution.

\subsection{Line defects with $SO(2)$ symmetry}
We now look at another holographic solution describing line defects with $SO(2)$ symmetry. In this case, we consider $F(4)$ gauged supergravity coupled to one vector multiplet with $SO(3)\sim SO(3)_R$ gauge group. The procedure is the same as in the previous case, so we will mainly give the results with less detail. By considering $SO(2)_R\subset SO(3)_R$ invariant scalars from $SO(4,1)/SO(4)$ coset described by the coset representative
\begin{equation}
L=e^{\phi_0Y_{01}+\phi_2Y_{31}}\label{SO2R_L}
\end{equation}
and using the gauge structure constants given in \eqref{SO2R_gauging}, we find the scalar potential
\begin{equation}
V=-e^{-6\sigma}\left[g_1^2e^{8\sigma}+4g_1me^{4\sigma}\cosh\phi_2-m^2\right]\label{V_SO2R}
\end{equation}
which admits only one supersymmetric $AdS_6$ vacuum at 
\begin{equation}
 \phi_2=0,\qquad \sigma=\frac{1}{4}\ln\left[\frac{3m}{g_1}\right],\qquad \ell^2=\frac{3\sqrt{3}}{4g_1\sqrt{g_1m}}\, . \label{AdS6_SO2R}  
\end{equation}    
As in the previous case, consistency of the BPS equations requires $\phi_0=0$. Therefore, we have set $\phi_0=0$ in the above equations for simplicity.
\\
\indent With the same ansatze for the remaining fields as in the previous case, we obtain the relevant BPS equations given in appendix \ref{detail}. In particular, we require either $\sin\theta=0$ or $\cos\theta=0$ for non-trivial scalars from the vector multiplet similar to the solutions considered in the previous analysis. With $\cos\theta=0$, we find the following BPS equations
\begin{eqnarray}
f'&=&h'=\frac{1}{2}e^{-3\sigma}(m+e^{4\sigma}g_1\cosh\phi_2),\\
\sigma&=&\frac{1}{2}\left(3me^{-3\sigma}-g_1e^\sigma\cosh\phi_2\right),\qquad Y'=\frac{1}{2}Yf',\\
b'&=&\frac{2e^{f-2\sigma}}{L},\qquad \phi'_2=-2g_1e^\sigma\sinh\phi_2
\end{eqnarray}    
with the solution given by
\begin{eqnarray}
& &f=h=-\frac{1}{3}\ln\sinh\phi_2+\frac{1}{12}\ln(\sigma_0\sinh\phi_2+3m\cosh\phi_2),\\
& &\sigma=\frac{1}{4}\ln\left[\frac{(\sigma_0+3m\coth\phi_2)\sinh\phi_2}{g_1}\right],\\
& &b=\frac{(\sigma_0+3m\coth\phi_2)^{\frac{1}{3}}}{mLg_1^{\frac{1}{4}}},\qquad Y=e^{\frac{f}{2}},\\
& &2g_1(\tilde{r}-\tilde{r}_0)=\ln\cosh\frac{\phi_2}{2}-\ln\sinh\frac{\phi_2}{2}
\end{eqnarray}
for a new radial coordinate $\tilde{r}$ defined by $\frac{d\tilde{r}}{dr}=e^\sigma$.
\\
\indent In this case, there is only one asymptotically locally supersymmetric $AdS_6$ vacuum at $\phi_2=\sigma=0$ for $g_1=3m$. It can also be seen that the above solution gives $\sigma\sim 0$ as $\phi_2\sim0$ for any values of the integration constant $\sigma_0$. The asymptotic form of the solution is given by
\begin{eqnarray}
\sigma\sim\phi_2\sim e^{-6mr}\sim z^3,\qquad f\sim -\ln z,\qquad b\sim z^{-1}
\end{eqnarray}
for $z=e^{-2mr}=e^{-\frac{r}{\ell}}$. However, unlike the previous solution, there is a singularity at a finite value of the radial coordinate $\tilde{r}$ at which the scalars diverge. There are a number of possibilities depending on the values of $\sigma_0$. We first consider when $\phi_2\rightarrow \infty$ for which the $\phi_2$ solution becomes
\begin{equation}
\phi_2\sim -\ln[g_1(\tilde{r}-\tilde{r}_0)].
\end{equation}  
For $\sigma_0=-3m$, we find that the solution near the singularity takes the form of
\begin{eqnarray}
& &\sigma\sim -\frac{1}{4}\phi_2\sim \frac{1}{4}\ln[g_1(\tilde{r}-\tilde{r}_0)],\qquad f\sim \frac{5}{12}\ln[g_1(\tilde{r}-\tilde{r}_0)],\nonumber \\
& & b\sim [g_1(\tilde{r}-\tilde{r}_0)]^{\frac{2}{3}}\rightarrow 0\, .
\end{eqnarray}  
For $\sigma_0\neq -3m$, we find
\begin{eqnarray}
& &\sigma\sim -\frac{1}{4}\ln[g_1(\tilde{r}-\tilde{r}_0)],\qquad f\sim \frac{1}{4}\ln[g_1(\tilde{r}-\tilde{r}_0)],\nonumber \\
& & b\sim (\sigma_0+3m)^{\frac{1}{3}}\sim\textrm{constant}\, .
\end{eqnarray}  
With the scalar potential given in \eqref{V_SO2R}, both of these behaviors lead to $V\rightarrow -\infty$. Accordingly, these singularities are physically acceptable by the criterion given in \cite{Gubser_singularity}. 
\\
\indent Other singular behaviors of the solution are given by $\phi_2\rightarrow -\infty$ with
\begin{equation}
\phi_2\sim \ln[g_1(\tilde{r}-\tilde{r}_0)].
\end{equation}
In this case, we find, for $\sigma_0=3m$, 
\begin{equation}
\sigma\sim \frac{1}{4}\ln[g_1(\tilde{r}-\tilde{r}_0)],\qquad f\sim \frac{5}{12}\ln[g_1(\tilde{r}-\tilde{r}_0)],\qquad b\rightarrow 0
\end{equation}
while for $\sigma_0\neq 3m$, the solution becomes
\begin{eqnarray}
& &\sigma\sim -\frac{1}{4}\ln[g_1(\tilde{r}-\tilde{r}_0)],\qquad f\sim \frac{1}{4}\ln[g_1(\tilde{r}-\tilde{r}_0)],\nonumber \\
& & b\sim (3m-\sigma_0)^{\frac{1}{3}}\sim\textrm{constant}\, .
\end{eqnarray}            
These are also acceptable since $V\rightarrow -\infty$ near the singularities.  
\\
\indent By the classification of supersymmetric $AdS_6$ vacua arising from consistent truncations of string/M-theory given in \cite{Henning_Malek_AdS7_6}, the matter-coupled $F(4)$ gauged supergravity coupled to one vector multiplet under consideration here can be embedded in type IIB theory via a consistent truncation on $S^2\times \Sigma$ with $\Sigma$ being a Riemann surface. We collect relevant formulae for uplifting the six-dimensional solution given above to ten dimensions in appendix \ref{truncation}. The resulting ten-dimensional fields are given by
\begin{eqnarray}
d\hat{s}^2&=&\frac{8c_6(3\rho)^{\frac{5}{4}}|dk|^{\frac{3}{2}}}{g_1^2\bar{\Delta}^{\frac{3}{4}}}\left[ \frac{g_1^2}{6\rho |dk|^2}\left[e^{2f}(ds^2_{AdS_2}+ds^2_{S^3})+e^{-2\sigma}d\tilde{r}^2\right] \right.\nonumber \\
& &+e^{2\sigma}\left(d\tilde{\theta}^2+\sin^2\tilde{\theta}d\tilde{\phi}^2+w^2-\frac{1}{\rho^2}p_ap_bn^an^b\right)+\frac{\widetilde{\Delta} e^{-2\sigma}}{\rho^2|dk|^2}dk^adp_a\nonumber \\
& &\left.+\frac{3}{\rho}e^{-2\sigma}n_a\cosh\phi_2\cos\tilde{\theta}dp^a \right],\\
H^{ab}&=&\frac{|dk|p^ap^be^{4\sigma}}{\sqrt{3\rho\bar{\Delta}}}+\sqrt{\frac{3\rho}{\bar{\Delta}}}\left(\sinh\phi_2\pd_ck^a\pd^cp^b+\cosh\phi_2\cos\tilde{\theta}\, n^{ac}\pd_cp^b\right),\\
C^a_{(2)}&=&-\frac{24c_6}{g_1^2}\left[(k^a+L^a)\textrm{vol}_{\tilde{S}^2}+\frac{|dk|^2}{\bar{\Delta}}\cosh\phi_2\sin^2\tilde{\theta}\sigma^a\wedge d\tilde{\phi}\right]\nonumber \\
& &-4c_6b(\tilde{r})p^a\textrm{vol}_{AdS_2},\\
F_{(5)}&=&F_{(2,3)}+F_{(3,2)}+*_{10}F_{(2,3)}
\end{eqnarray}
with the $2\times 2$ matrix $H^{ab}$ describing the ten-dimensional dilaton $\Phi$ and axion $C_0$ as shown in \eqref{Hab}. Various form fields appearing in the self-dual $5$-form field strength are defined by
\begin{eqnarray}
F_{(3,2)}&=&-\frac{288c_6^2|dk|}{g_1^2}\left\{\frac{\rho |dk|}{\bar{\Delta}}b'(\tilde{r})d\tilde{r}\wedge \textrm{vol}_{AdS_2}\wedge\left[(\sinh^2\phi_2-\cosh^2\phi_2\cos^2\tilde{\theta})\textrm{vol}_{\tilde{S}^2}\right.\right.\nonumber \\
& &\left.\phantom{\frac{|d|}{\bar{\Delta}}} \left.-\cosh\phi_2\sin^2\tilde{\theta}w\wedge d\tilde{\phi} \right]+b'(\tilde{r})e^{f-4\sigma}\textrm{vol}_{S^3}\wedge \textrm{vol}_\Sigma \right\},\nonumber \\
F_{(2,3)}&=&\frac{96c_6^2|dk|^2}{g_1^2\bar{\Delta}}e^{4\sigma}\textrm{vol}_{AdS_2}\wedge \left[\rho^ap_a\wedge \textrm{vol}_{\tilde{S}^2} \right.\nonumber \\
& &\left. -p_ap_bn^{ac}\pd_ck^b\cosh\phi_2\sin^2\tilde{\theta}d\tilde{\theta}\wedge \textrm{vol}_\Sigma\right]
\end{eqnarray}
with $F_{(3,2)}=*_{10}F_{(3,2)}$. Various quantities appearing in these equations are defined in appendix \ref{truncation}. For clarity, we have explicitly shown the argument of the function $b(\tilde{r})$ in the six-dimensional two-form field.
\\
\indent From the ten-dimensional metric $d\hat{s}^2$, we find that 
\begin{equation}
\hat{g}_{00}=-\frac{4c_6(3\rho\bar{\Delta})^{\frac{1}{4}}e^{2f}}{|dk|^{\frac{1}{2}}}\, .
\end{equation}
For $\phi_2\rightarrow \infty$, we find 
\begin{eqnarray}
\hat{g}_{00}\sim e^{-\frac{1}{3}\phi_2}\rightarrow 0\qquad 
\textrm{and}\qquad \hat{g}_{00}\sim \textrm{constant}
\end{eqnarray}
for $\sigma_0=-3m$ and $\sigma_0\neq -3m$, respectively. Similarly, as $\phi_2\rightarrow -\infty$, the six-dimensional solution gives 
\begin{eqnarray}
\hat{g}_{00}\sim e^{\frac{1}{3}\phi_2}\rightarrow 0\qquad 
\textrm{and}\qquad \hat{g}_{00}\sim \textrm{constant}
\end{eqnarray}
for $\sigma_0=3m$ and $\sigma_0\neq 3m$, respectively. These results imply that all the singular behaviors in the six-dimensional solution are physically acceptable in type IIB theory by the criterion of \cite{maldacena_nogo}. Consequently, the solution should holographically describe a conformal line defect within five-dimensional $N=2$ SCFT dual to the $AdS_6$ vacuum \eqref{AdS6_SO2R}.
%%%%%%%%%%%%%%%%%%%%%%%%%%%%%%%%%%%%%%%%%%%%%%%%%%%%%%%%%%%%%%%%%%%%%%%%%%%%%%%%%%%%%%%%%%%%%%%%%%%%%%%%%%%%%%%%%%%%%%%%%%%%%%%%%%%%%%%%%
\section{Surface defects from matter-coupled $F(4)$ gauged supergravity}\label{surface_defects}
We now move to another type of solutions describing two-dimensional conformal defects within five-dimensional $N=2$ SCFTs. Following \cite{surface_defect_F4_Dibitetto}, we use the metric ansatz of the form
\begin{equation}
ds^2=e^{2f(r)}dx^2_{AdS_3}+dr^2+e^{2h(r)}ds^2_{S^2}\, .\label{surface_metric}
\end{equation}
This takes a similar form to the case of line defects, but the slicing is now given by $AdS_3\times S^2$. The solutions then preserve two-dimensional conformal symmetry leading to a holographic description of conformal surface defects within $N=2$ SCFTs in five dimensions.
\\
\indent The six-dimensional coordinates in this case are split as $x^\mu=(x^\alpha,r,\theta^i)$ with $\alpha=0,1,2$ and $i=1,2$. To support the curvature of the slicing, we again turn on the two-form field of the form
\begin{equation}
B=b(r)\textrm{vol}_{S^2}\, .
\end{equation}
\indent The Killing spinors are taken to be 
\begin{equation}
\epsilon=Y\eta\otimes\left[\cos\theta\chi^+\otimes \epsilon_0+i\sin\theta\gamma_*\chi^+\otimes \sigma_3\epsilon_0-i\cos\theta\gamma_*\chi^-\otimes \sigma_3\epsilon_0+\sin\theta\chi^-\epsilon_0\right]
\end{equation}
in which the Killing spinors $\eta$ and $\chi^\pm$ on $AdS_3$ and $S^2$ respectively satisfy
\begin{equation}
\hat{\nabla}_\alpha=\frac{1}{2L}\rho_\alpha \eta\qquad \textrm{and}\qquad \tilde{\nabla}_i\chi^\pm=\frac{1}{2R}\gamma_*\tilde{\gamma}_i\chi^\mp\, .\label{Killing_eq_surface}
\end{equation}
$Y$ and $\theta$ are functions of $r$ as in the previous section. In this case, the explicit representation of gamma matrices is chosen to be
\begin{eqnarray}
& &\gamma^\alpha=\rho^\alpha\otimes \mathbb{I}_2\otimes \sigma_2,\qquad \gamma^r=-(\mathbb{I}_2\otimes \gamma_*\otimes \sigma_1),\nonumber \\
& &\gamma^{i}=-(\mathbb{I}_2\otimes \tilde{\gamma}^i\otimes \sigma_1), \qquad \gamma_7=i\mathbb{I}_2\otimes \mathbb{I}_2\otimes \sigma_3
\end{eqnarray}
with $\gamma_*=i\tilde{\gamma}^1\tilde{\gamma}^2$ and $\gamma_*^2=\mathbb{I}_2$.
\\
\indent We now impose the following projection condition 
\begin{equation}
(\gamma_*\otimes \sigma_1)(\chi^\pm\otimes \epsilon_0)=\pm \chi^\pm \otimes \epsilon_0
\end{equation}
which is equivalent to 
\begin{equation}
\chi^\pm\otimes \sigma_2\epsilon_0=\mp i\gamma_* \chi^\pm \otimes \sigma_3\epsilon_0
\end{equation}
and find the BPS equations as in the case of line defect solutions. The detail on this is given in appendix \ref{detail}. The relevant analysis also closely follows that performed in the previous section, so we will only give the main results with less detail to avoid repetition. In particualr, in all the following solutions, the gaugino variations restrict the phase of the Killing spinors to satisfy the conditions $\sin\theta=0$ or $\cos\theta=0$ as in the previous section. 

\subsection{Surface defects with $SO(3)_{\textrm{diag}}$ symmetry}
We begin with the solution interpolating between two supersymmetric $AdS_6$ vacua with an $SO(3)_{\textrm{diag}}$ singlet scalar given in \eqref{SO3diag_L}. We will set $\sin\theta=0$ which again gives $h(r)=f(r)$ and the following BPS equations
\begin{eqnarray}
f'&=&\frac{1}{2}e^{-3\sigma}\left[m+e^{4\sigma}(g_1\cosh^3\phi-g_2\sinh^3\phi)\right],\\
\sigma'&=&\frac{1}{2}e^{-3\sigma}\left[3m-e^{4\sigma}(g_1\cosh^3\phi-g_2\sinh^3\phi)\right],\\
b'&=&-\frac{4e^{f-2\sigma}}{L},\qquad Y'=\frac{1}{2}Yf',\\
\phi'&=&-e^\sigma\sinh2\phi(g_1\cosh\phi-g_2\sinh\phi)
\end{eqnarray}
together with two algebraic constraints
\begin{equation}
b=-\frac{2e^{f+\sigma}}{Lm}\qquad \textrm{and}\qquad R=2L\, .
\end{equation}
As in the previous section, we have verified that the resulting BPS equations are compatible with these constraints and the corresponding field equations for all the fields. By the same procedure as in the previous section, we find the solution 
\begin{eqnarray}
f&=&\frac{1}{4}\ln(g_1\cosh\phi-g_2\sinh\phi)-\frac{1}{3}\ln\sinh2\phi \nonumber \\
& &+\frac{1}{12}\left[6m\cosh2\phi-\frac{3m(g_1^2+g_2^2)\sinh2\phi}{g_1g_2}\right],\nonumber \\
\sigma&=&\frac{1}{4}\ln\left[\frac{3m\cosh\phi}{g_1}-\frac{3m\sinh\phi}{g_2}\right],\\
b&=&-\frac{2^{\frac{3}{4}}3^{\frac{1}{3}}}{Lm^{\frac{2}{3}}}\left[\frac{g_1^2+g_2^2-2g_1g_2\coth\phi-g_1g_2\tanh\phi}{g_1g_2m^2}\right]^{\frac{1}{3}},\\
Y&=&e^{\frac{f}{2}},\nonumber \\
2g_1g_2(\tilde{r}-\tilde{r}_0)&=&g_2\ln\cosh\frac{\phi}{2}-g_2\ln\sinh\frac{\phi}{2}-2g_1\tan^{-1}\tanh\frac{\phi}{2}\nonumber \\
& &-2\sqrt{g_2^2-g_1^2}\tanh^{-1}\left[\frac{g_2g_1\tanh\frac{\phi}{2}}{\sqrt{g_2^2-g_1^2}}\right]
\end{eqnarray}
after choosing an appropriate integration constant for $\sigma$. This solution is the same up to a numerical factor in $b$ as in the case of line defects. All the asymptotic behaviors are essentially the same, so we will not repeat these again. In particular, we just note that there is a source of a dimension-$4$ operator dual to the two form field turned on along the flow.

\subsection{Surface defects with $SO(2)$ symmetry}
Similarly, we can look for solutions holographically dual to surface defects with $SO(2)_R\subset SO(3)_R$ symmetry for $F(4)$ gauged supergravity coupled to one vector multiplet with $SO(3)_R\times U(1)$ gauge group. Repeating the same analysis leads to the BPS equations
\begin{eqnarray}
f'&=&\frac{1}{2}\left[e^{-3\sigma}m+g_1e^\sigma\cosh\phi_2\right],\\
\sigma'&=&\frac{1}{2}\left[3me^{-3\sigma}-g_1e^\sigma \cosh\phi_2\right],\\
b'&=&-\frac{4e^{f-2\sigma}}{L},\qquad Y'=\frac{1}{2}Yf',\\
\phi_2'&=&-2g_1e^\sigma\sinh\phi_2 
\end{eqnarray} 
and
\begin{equation}
b=-\frac{2e^{f+\sigma}}{Lm}\qquad \textrm{and}\qquad L=2R\, .
\end{equation}
As in the previous solution, the solution is the same as that of the line defects up to a numerical factor in the two-form field. All the possible singularities are also the same as well as the uplifted solutions in type IIB theory with the six-dimensional metric replaced by \eqref{surface_metric} and $(\textrm{vol}_{AdS_2},\textrm{vol}_{S^3})$ by $(\textrm{vol}_{AdS_3},\textrm{vol}_{S^2})$. We will not repeat all these results again.             

%%%%%%%%%%%%%%%%%%%%%%%%%%%%%%%%%%%%%%%%%%%%%%%%%%%%%%%%%%%%%%%%%%%%%%%%%%%%%%%%%%%%%%%%%%%%%%%%%%%%%%%%%%%%%%%%%%%%%%%%%%%%%%%%%%%%%%%%%
\section{Conclusions}\label{conclusion}
We have studied solutions of matter-coupled $F(4)$ gauged supergravity in the form of $AdS_2\times S^3$- and $AdS_3\times S^2$-sliced domain walls. Unlike the solutions considered in \cite{AdS2_F4_Dibitetto} and \cite{surface_defect_F4_Dibitetto}, the presence of scalar fields from the vector multiplets imposes a condition that constrains the warp factors of $AdS_2$ ($AdS_3$) and $S^3$ ($S^2$) to be equal and fixes the phase function of the Killing spinors to be constant. Therefore, the resulting solutions take the form of ``charged'' domain walls as pointed out in \cite{AdS2_F4_Dibitetto} and \cite{surface_defect_F4_Dibitetto}. This is very similar to the solution describing surface defects in $N=(1,0)$ SCFT in six dimensions found in matter-coupled seven-dimensional $N=2$ gauged supergravity studied in \cite{7D_N2_DW_3_form}. This could arise from the fact that in both cases, the scalars from vector multiplets do not couple to the two-form and three-form fields from the supergravity multiplet. It would be interesting to find the field theory description of this result.
\\
\indent We have considered two $F(4)$ gauged supergravities coupled to three and one vector multiplets with $SO(3)_R\times SO(3)$ and $SO(3)_R\times U(1)$ gauge groups. The resulting solutions should holographically describe conformal line and surface defects within $N=2$ SCFTs in five dimensions with a source of a dimension-$4$ operator dual to the two-form field turned on along the flows between two conformal fixed points and flows from a conformal fixed point to non-conformal phases. In the case of $SO(3)_R\times U(1)$ gauge group, the higher-dimensional origin as a consistent truncation of type IIB theory on $S^2\times \Sigma$ is known. We have also given the corresponding ten-dimensional solutions by uplifting the six-dimensional solutions of $F(4)$ gauged supergravity.      
\\
\indent It would be interesting to identify line and surface defects, in terms of position dependent deformations, dual to the solutions found here in the framework of five-dimensional SCFTs. It is also interesting to use holographic renormalization to compute the expectation values of the defect operators and boundary stress tensor as done in \cite{line_defect_F4_Gutperle} for the case of line defect solution found in \cite{AdS2_F4_Dibitetto} and match with the field theory results. Finally, finding type IIB solutions describing possible brane configurations, of which the uplifted solutions given here describe low-energy regimes as in the case of pure $F(4)$ gauged supergravity studied in \cite{AdS2_F4_Dibitetto} and \cite{surface_defect_F4_Dibitetto}, is also worth considering. We note that although the six-dimensional solutions for both line and surface defects are the same, the uplifted solutions could correspond to different brane configurations in type IIB theory. In particular, the uplifted solution for line defects would involve F1-D1-branes, due to the presence of two-form fluxes along $AdS_2$, and D3-branes corresponding to a non-trivial four-form field on top of the 5-brane web dual to the SCFT in five dimensions \cite{IIB_AdS6,04_defect_Nicolo}. On the other hand, we expect the brane configuration corresponding to the surface defects to involve only D3-branes without F1-D1-branes since the electric part of the two-form field does not appear in the uplifted ten-dimensional solution.               
\vspace{0.5cm}\\
%%%%%%%%%%%%%%%%%%%%%%%%%%%%%%%%%%%%%%%%%%%%%%%%%%%%%%%%%%%%%%%%%%%%%%%%%%%%%%%%%%%%%%%%%%%%%%%%%%%%%%%%%%%%%%%%%%%%%%%%%%%%%%%%%%%%%%%%%
{\large{\textbf{Acknowledgement}}} \\
This work is funded by National Research Council of Thailand (NRCT) and Chulalongkorn University under grant N42A650263. The author would like to thank Emanuel Malek and Christoph F. Uhlemann for helpful correspondences. 
%%%%%%%%%%%%%%%%%%%%%%%%%%%%%%%%%%%%%%%%%%%%%%%%%%%%%%%%%%%%%%%%%%%%%%%%%%%%%%%%%%%%%%%%%%%%%%%%%%%%%%%%%%%%%%%%%%%%%%%%%%%%%%%%%%%%%%%%%
\appendix
\section{Derivation of the BPS equations}\label{detail}
In this appendix, we give more detail on the derivation of the relevant BPS equations for obtaining solutions of line and surface defects. The $SO(2)_R$ sector in the case of one vector multiplet with $SU(2)_R\times U(1)$ gauge group is effectively the same as $SO(2)_R\times SO(2)$ invariant sector of $F(4)$ gauged supergravity coupled to three vector multiplets. Furthermore, both the $SO(3)_{\textrm{diag}}$ and $SO(2)_R\times SO(2)$ sectors in the latter can be obtained from the $SO(2)_{\textrm{diag}}\subset SO(2)_R\times SO(2)$ invariant sector. We will accordingly give the BPS equations in this $SO(2)_{\textrm{diag}}$ sector and subsequently choose suitable subtruncations of interest in each case.
\\
\indent As in \cite{AdS6_BH} and \cite{6D_Janus_RG}, the coset representative in the $SO(2)_{\textrm{diag}}$ sector is given by
\begin{equation}
L=e^{\phi_0 Y_{03}}e^{\phi_1 (Y_{11}+Y_{22})}e^{\phi_2 Y_{33}}e^{\phi_3 (Y_{12}-Y_{21})}\, .\label{L_SO2diag}
\end{equation}
We also need to set $\phi_3=0$ in order to consistently truncate out all the gauge fields as pointed out in \cite{6D_Janus_RG}. To obtain solutions for $SO(3)_{\textrm{diag}}$ singlet scalars, we simply set $\phi_0=0$ and $\phi_2=\phi_1=\phi$ while the solutions with $SO(2)_R$ symmetry in the case of one vector multiplet and $SU(2)_R\times U(1)$ gauge group are recovered by setting $\phi_1=0$.     

\subsection{Line defects}
From the metric \eqref{line_metric}, we will choose the following choice of vielbein
\begin{equation}
e^{\hat{\alpha}}=e^f\hat{e}^{\hat{\alpha}},\qquad e^{\hat{r}}=dr,\qquad e^{\hat{i}}=e^h\tilde{e}^{\hat{i}}
\end{equation}
with $\hat{e}^{\hat{\alpha}}$ and $\tilde{e}^{\hat{i}}$ being the vielbeins on $AdS_2$ and $S^3$, respectively. We then find non-vanishing components of the spin connection
\begin{eqnarray}
{\omega^{\hat{\alpha}}}_{\hat{\beta}}={\hat{\omega}^{\hat{\alpha}}}_{\phantom{\hat{\alpha}}\hat{\beta}},\qquad {\omega^{\hat{\alpha}}}_{\hat{r}}=f'e^{\hat{\alpha}},\qquad {\omega^{\hat{i}}}_{\hat{j}}={\tilde{\omega}^{\hat{i}}}_{\phantom{\hat{i}}\hat{j}}, \qquad {\omega^{\hat{i}}}_{\hat{r}}=h'e^{\hat{i}} 
\end{eqnarray}
with ${\hat{\omega}^{\hat{\alpha}}}_{\phantom{\hat{\alpha}}\hat{\beta}}$ and ${\tilde{\omega}^{\hat{i}}}_{\phantom{\hat{i}}\hat{j}}$ being the spin connections on $AdS_2$ and $S^3$, respectively.
\\
\indent For convenience, we recall that the ansatz for the Killing spinors is given by
\begin{equation}
\epsilon=Y\left[i\cos\theta\chi^+\otimes \epsilon_0+i\sin\theta\chi^+\otimes \sigma_3\epsilon_0+\sin\theta\chi^-\otimes \epsilon_0-\cos\theta\chi^-\otimes \sigma_3\epsilon\right]\otimes \eta
\end{equation}
subject to the projection condition
\begin{equation}
\sigma_2\epsilon_0=\epsilon_0\, .
\end{equation}
We also note that this projector implies 
\begin{equation}
\sigma_1\epsilon_0=i\sigma_3\epsilon_0
\end{equation}
which is a useful relation in the following computation. 
\\
\indent To find the BPS equations, we use the ansatze for all the bosonic fields given in section \ref{line_defects} together with the above projectors in the supersymmetry transformations of fermions and separately solve the resulting conditions for the four components $\chi^\pm\otimes \epsilon_0\otimes \eta$ and $\chi^\pm\otimes \sigma_3\epsilon_0\otimes \eta$. We first consider $\delta\chi^A=0$ equations. The $\chi^+\otimes \epsilon_0\otimes \eta$ and $\chi^-\otimes \sigma_3\epsilon_0\otimes \eta$ components give the same condition of the form
\begin{equation}
\left[\frac{1}{2}\sigma'+(N_0-iN_3)\right]\cos\theta+\left[\frac{1}{8}me^{-\sigma-2f}{(L^{-1})^0}_0b-\frac{1}{16}e^{2\sigma-2f}b'\right]\sin\theta=0\label{eq1}
\end{equation}
while for $\chi^+\otimes \sigma_3\epsilon_0\otimes \eta$ and $\chi^-\otimes \epsilon_0\otimes \eta$ components, we find
\begin{equation}
\left[-\frac{1}{2}\sigma'+(N_0-iN_3)\right]\sin\theta+\left[\frac{1}{8}me^{-\sigma-2f}{(L^{-1})^0}_0b+\frac{1}{16}e^{2\sigma-2f}b'\right]\cos\theta=0\, .\label{eq2}
\end{equation}
In these equations, we have defined the following quantities
\begin{eqnarray}
N_0&=&\frac{1}{8}e^{-3\sigma}\cosh\phi_0\left[6m+g_2e^{4\sigma}\sinh\phi_2(\cosh2\phi_1-1)\right]\nonumber \\
& &-\frac{1}{8}g_1e^\sigma\cosh\phi_2(\cosh2\phi_1+1),\nonumber \\
N_3&=&\frac{1}{8}e^{-3\sigma}\sinh\phi_0\left[e^{4\sigma}g_2(\cosh2\phi_1-1)-6m\sinh\phi_2\right]\label{N0_def}
\end{eqnarray}
and ${(L^{-1})^0}_0=\cosh\phi_0$. From these equations, we immediately see that consistency requires $N_3=0$ which in turns leads to $\phi_0=0$. Accordingly, we will set $\phi_0=0$ and ${(L^{-1})^0}_0=1$ from now on. 
\\
\indent We then move to the gravitino variations $\delta\psi^A_r$ which give
\begin{equation}
\left[Y'-S_0Y\right]\cos\theta+\left[Y\theta'+\frac{1}{8}me^{-\sigma-2f}bY-\frac{1}{16}e^{2\sigma-2f}b'Y\right]\sin\theta=0\label{eq3}
\end{equation}
for $\chi^+\otimes \epsilon_0\otimes \eta$ and $\chi^-\otimes \sigma_3\epsilon_0\otimes \eta$ components and
\begin{equation}
\left[Y'+S_0Y\right]\sin\theta+\left[Y\theta'+\frac{1}{8}me^{-\sigma-2f}bY+\frac{1}{16}e^{2\sigma-2f}b'Y\right]\cos\theta=0\label{eq4}
\end{equation}
for $\chi^+\otimes \sigma_3\epsilon_0\otimes \eta$ and $\chi^-\otimes \epsilon_0\otimes \eta$ components. In these equations, $S_0$ is given by
\begin{eqnarray}
S_0&=&-\frac{1}{8}g_1e^\sigma\cosh\phi_2(\cosh2\phi_1+1)\nonumber \\
& &-\frac{1}{8}e^{-3\sigma}\left[2m-g_2e^{4\sigma}\sinh\phi_2(\cosh2\phi_1-1)\right].\label{S0_def}
\end{eqnarray}
\indent Taking into account the Killing spinor equations on $AdS_2$ satisfying \eqref{AdS2_S3_Killing}, we find the following BPS equations from $\delta\psi^A_\alpha$ 
\begin{eqnarray}
\left[\frac{1}{2}f'-S_0\right]\cos\theta-\left[\frac{1}{2L}e^{-f}-\frac{3}{8}e^{-\sigma-2f}b-\frac{1}{16}e^{2\sigma-2f}b'\right]\sin\theta=0,& &\quad \label{eq5}\\
\left[\frac{1}{2}f'+S_0\right]\sin\theta+\left[\frac{1}{2L}e^{-f} -\frac{3}{8}e^{-\sigma-2f}b+\frac{1}{16}e^{2\sigma-2f}b'\right]\cos\theta=0,& &\quad \label{eq6}
\end{eqnarray}
for $(\chi^+\otimes \epsilon_0\otimes \eta,\chi^-\otimes \sigma_3\epsilon_0\otimes \eta)$ and $(\chi^+\otimes \sigma_3\epsilon_0\otimes \eta,\chi^-\otimes \epsilon_0\otimes \eta)$ components, respectively
\\
\indent Similarly, $\delta\psi^A_i$ conditions give
\begin{eqnarray}
\left[\frac{1}{2}h' +S_0\right]\sin\theta-\left[\frac{1}{2R}e^{-h}-\frac{1}{8}e^{-\sigma-2f}b+\frac{1}{16}e^{2\sigma-2f}b'\right]\cos\theta=0,& &\qquad\label{eq7}\\
\left[\frac{1}{2}h'-S_0\right]\cos\theta-\left[\frac{1}{2R}e^{-h}+\frac{1}{8}e^{-\sigma-2f}b+\frac{1}{16}e^{2\sigma-2f}b'\right]\sin\theta=0,& &\qquad \label{eq8}
\end{eqnarray}
\indent Finally, we consider the gaugino variations $\delta\lambda^{AI}$. From $\delta\lambda^{1A}$ and $\delta\lambda^{2A}$, we obtain the followng conditions
\begin{eqnarray}
(\phi'_1+M_1)\cos\theta=0\qquad \textrm{and}\qquad (\phi'_1-M_1)\sin\theta=0 
\end{eqnarray}
while $\delta\lambda^{3A}$ gives
\begin{eqnarray}
(\phi'_2+M_3)\cos\theta=0\qquad \textrm{and}\qquad (\phi'_2-M_3)\sin\theta=0 
\end{eqnarray}
with 
\begin{eqnarray}
& &M_1=-e^\sigma\sinh2\phi_1(g_1\cosh\phi_2-g_2\sinh\phi_2),\nonumber \\
& &M_3=g_2e^\sigma\cosh\phi_2(\cosh2\phi_1-1)-g_1e^\sigma\sinh\phi_2(\cosh2\phi_1+1).\label{M_def}
\end{eqnarray}
\indent We can now solve equations \eqref{eq1} to \eqref{eq8} and find the following BPS equations
\begin{eqnarray}
f'&=&\frac{1}{4L}\left[e^{-f}\tan2\theta+L(3N_0+5S_0-3(N_0-S_0)\cos4\theta)\sec2\theta\right],\\
h'&=&\frac{1}{2L}\sec2\theta\left[e^{-f}\cos\theta\sin\theta+4S_0L+(S_0-N_0)L\sin^22\theta\right],\\
b'&=&-\frac{2}{L}e^{f-2\sigma}\sec2\theta\left[1+2e^fL(S_0+3N_0)\sin2\theta\right],\\
\sigma'&=&\frac{1}{2L}\sec2\theta\left[L(N_0-S_0)\sin^22\theta-4N_0L-e^{-f}\cos\theta\sin\theta\right],\\
Y'&=&\frac{Y}{8L}e^{-f}\sec2\theta\left[Le^f(5S_0+3N_0)+\sin2\theta-3Le^f(N_0-S_0)\cos4\theta\right],\quad \\
\theta'&=&(N_0-S_0)\sin2\theta
\end{eqnarray}
with two algebraic constraints
\begin{eqnarray}
b&=&\frac{e^{f+\sigma}}{L m}\left[1+2e^fL(S_0-N_0)\sin2\theta\right],\\
R&=&\frac{1}{2}e^h\left[Le^{-f}\sec 2\theta+2(3S_0+N_0)\tan2\theta\right].
\end{eqnarray}
These constraints arise from the fact that there are more equations than unknown functions. In particular, equations \eqref{eq1} to \eqref{eq8} lead to three different expressions for $b'$. Consequently, consistency among these leads to two algebraic constraints. We also note that these equations reduce to those considered in \cite{AdS2_F4_Dibitetto} for $\phi_1=\phi_2=0$ up to some redefinition of variables.

\subsection{Surface defects}
In the case of surface defects with an $AdS_3\times S^2$-sliced metric given by \eqref{surface_metric}, we use the following vielbein
\begin{equation}
e^{\hat{\alpha}}=e^f\hat{e}^{\hat{\alpha}},\qquad e^{\hat{r}}=dr,\qquad e^{\hat{i}}=e^h\tilde{e}^{\hat{i}} 
\end{equation} 
and the spin connection   
\begin{eqnarray}
{\omega^{\hat{\alpha}}}_{\hat{\beta}}={\hat{\omega}^{\hat{\alpha}}}_{\phantom{\hat{\alpha}}\hat{\beta}},\qquad {\omega^{\hat{\alpha}}}_{\hat{r}}=f'e^{\hat{\alpha}},\qquad{\omega^{\hat{i}}}_{\hat{j}}={\tilde{\omega}^{\hat{i}}}_{\phantom{\hat{i}}\hat{j}},\qquad {\omega^{\hat{i}}}_{\hat{r}}=h'e^{\hat{i}}
\end{eqnarray}
In this case, $(\hat{e}^{\hat{\alpha}},\tilde{e}^{\hat{i}})$ and $({\hat{\omega}^{\hat{\alpha}}}_{\phantom{\hat{\alpha}}\hat{\beta}},{\tilde{\omega}^{\hat{i}}}_{\phantom{\hat{i}}\hat{j}})$ are respectively the corresponding vielbeins and spin connections on $AdS_3$ and $S^2$ with $\alpha=0,1,2$ and $i=1,2$.
\\
\indent The Killing spinors are of the form
\begin{equation}
\epsilon=Y\eta\otimes \left[\cos\theta\chi^+\otimes \epsilon_0+i\sin\theta\gamma_*\chi^+\otimes \sigma_3\epsilon_0-i\cos\theta\gamma_*\chi^-\otimes \sigma_3\epsilon_0+\sin\theta\chi^-\otimes \epsilon_0\right]
\end{equation}
with $\epsilon_0$ being a two-component constant spinor. $\eta$ and $\chi^\pm$ are respectively Killing spinors on $AdS_3$ and $S^2$ satisfying the conditions given in \eqref{Killing_eq_surface}. By imposing the projection condition
\begin{equation}
(\gamma_*\otimes \sigma_1)(\chi^\pm\otimes \epsilon_0)=\pm \chi^\pm\otimes \epsilon_0
\end{equation}
which also implies
\begin{equation}
\chi^\pm\otimes \sigma_2\epsilon_0=\mp i\gamma_*\chi^\pm\otimes \sigma_3\epsilon_0,
\end{equation}
we can determine all the BPS equations by the same procedure as in the case of line defects shown previously. In this case, we separately solve the resulting conditions for the four components $\eta\otimes \chi^\pm\otimes \epsilon_0$ and $\eta\otimes \gamma_*\chi^\pm\otimes \sigma_3\epsilon_0$. As in the previous case, it turns out that consistency requires $\phi_0=0$. 
\\
\indent The conditions obtained from $\delta\chi^A$ and $\delta\psi^A_\mu$ are given by
\begin{eqnarray}
\left[\frac{1}{2}\sigma'-N_0\right]\cos\theta+\left[\frac{1}{8}me^{-\sigma-2h}+\frac{1}{16}e^{2\sigma-2h}b'\right]\sin\theta=0, & &\quad\\
\left[\frac{1}{2}\sigma'+N_0\right]\sin\theta+\left[\frac{1}{8}me^{-\sigma-2h}b-\frac{1}{16}e^{2\sigma-2h}b'\right]\cos\theta=0,& &\\
\left[Y'+S_0Y\right]\cos\theta-\left[Y\theta'+\frac{1}{8}me^{-\sigma-2h}bY+\frac{1}{16}e^{2\sigma-2h}Yb'\right]\sin\theta=0, & &\\
\left[Y'-S_0Y\right]\sin\theta+\left[Y\theta'-\frac{1}{8}me^{-\sigma-2h}bY+\frac{1}{16}e^{2\sigma-2h}Yb'\right]\cos\theta=0,& &\\
\left[\frac{1}{2}f'+S_0\right]\cos\theta+\left[\frac{1}{2L}e^{-f}-\frac{1}{8}me^{-\sigma-2h}b+\frac{1}{16}e^{2\sigma-2h}b'\right]\sin\theta=0,& &\\
\left[\frac{1}{2}f'-S_0\right]\sin\theta-\left[\frac{1}{2L}e^{-f}+\frac{1}{8}me^{-\sigma-2h}b+\frac{1}{16}e^{2\sigma-2h}b'\right]\cos\theta=0, & &\\
\left[\frac{1}{2}h'+S_0\right]\cos\theta+\left[\frac{1}{2R}e^{-h}+\frac{3}{8}me^{-\sigma-2h}b-\frac{1}{16}e^{2\sigma-2h}b'\right]\sin\theta=0, & &\\
\left[\frac{1}{2}h'-S_0\right]\sin\theta+\left[\frac{1}{2R}e^{-h}+\frac{3}{8}me^{-\sigma-2h}b+\frac{1}{16}e^{2\sigma-2h}b'\right]\cos\theta=0\, . & &
\end{eqnarray}
$N_0$ and $S_0$ are the same as in the previous case and given in \eqref{N0_def} and \eqref{S0_def}. These equations can be solved and lead to the following BPS equations  
\begin{eqnarray}
f'&=&-\frac{1}{2L}\left[e^{-f}\tan2\theta+4LS_0\cos2\theta+2L(N_0+S_0)\sin2\theta\tan2\theta\right], \\
h'&=&-\frac{1}{2L}e^{-f}\sec2\theta\left[\sin2\theta+e^fL(N_0-S_0)\cos4\theta-e^fL(N_0-5S_0)\right],\quad\\
b'&=&-\frac{1}{L}e^{-f+2h-2\sigma}\left[4-8Le^f(N_0-S_0)\sin2\theta\right],\\
\sigma'&=&\frac{1}{2L}\left[\left(e^{-f}+2L(N_0+S_0)\sin2\theta\right)\tan2\theta+4LN_0\cos2\theta\right],\\
Y'&=&-\frac{1}{4L}\left[e^{-f}Y\tan2\theta\left(1+2Le^f(N_0+S_0)\sin2\theta\right)+4LS_0Y\cos2\theta\right],\\
\theta'&=&(S_0-N_0)\sin2\theta
\end{eqnarray}
together with two algebraic constraints
\begin{eqnarray} 
b&=&-\frac{2}{Lm}e^{\sigma-f+2h}\sec2\theta\left[1+2e^fL(N_0+S_0)\sin2\theta\right],\\
R&=&\frac{e^{f-2h}L\cos2\theta}{2+2Le^f(N_0+3S_0)\sin2\theta}\, .
\end{eqnarray}
As in the previous case, for $\phi_1=\phi_2=0$, we recover the BPS equations studied in \cite{surface_defect_F4_Dibitetto} for pure $F(4)$ gauged supergravity.
\\
\indent Finally, the variations $\delta\lambda^{AI}$ give
\begin{eqnarray}
& &(\phi'_1-M_1)\cos\theta=0,\qquad (\phi'_1+M_1)\sin\theta=0,\\
& &(\phi'_2+M_3)\cos\theta=0,\qquad (\phi'_2-M_3)\sin\theta=0
\end{eqnarray}
with $M_1$ and $M_3$ given in \eqref{M_def}.

%%%%%%%%%%%%%%%%%%%%%%%%%%%%%%%%%%%%%%%%%%%%%%%%%%%%%%%%%%%%%%%%%%%%%
\section{Consistent truncation of type IIB theory on $S^2\times \Sigma$}\label{truncation}
In this appendix, we collect relevant formulae for uplifting solutions of six-dimensional $F(4)$ gauged supergravity coupled to one vector multiplet to type IIB theory as given in \cite{Henning_Malek_AdS7_6}. In particular, we mainly consider $SO(2)_R$ singlet sector of the $F(4)$ gauged supergravity with two scalars denoted by $\phi_0$ and $\phi_2$ and the two-form field. Although the solutions found in this paper only involve non-vanishing $\phi_2$, we give all the formulae of the truncation for both $\phi_0$ and $\phi_2$ non-vanishing. The result can be useful in other applications such as uplifting the Janus solution found in \cite{6D_Janus} by setting the two-form field to zero. 
\\
\indent Consistent truncations of type IIB theory on a product of a two-sphere and a Riemann surface $S^2\times \Sigma$ are classified by two holomorphic functions $f^a$, $a=1,2$, written as
\begin{equation}
f^a=-p^a+ik^a\, .
\end{equation}
These holomorphic functions must satisfy the following consistency conditions
\begin{equation}
i\pd f^a \bar{\pd}\bar{f}_a\geq 0\qquad \textrm{and}\qquad \rho\geq 0 
\end{equation}
with $\rho$ defined via
\begin{equation}
d\rho=-p_adk^a\, .
\end{equation}
Indices $a,b$ can be raised and lowered by the convention $p_a=p^b\epsilon_{ba}$ and $k^a=\epsilon^{ab}k_b$. To avoid confusion with the $S^2$ inside the six-dimensional metric of the solutions for surface defects, we will denote the two-sphere in $S^2\times \Sigma$ by $\tilde{S}^2$. Coordinates on $\Sigma$ and $\tilde{S}^2$ are denoted respectively by $x^a=(x^1,x^2)$ and $y^i$, $i=1,2,3$, satisfying $y^iy^j\delta_{ij}=1$. The latter can be chosen to be
\begin{equation}
y^1=\sin\tilde{\theta}\cos\tilde{\varphi},\qquad y^2=\sin\tilde{\theta}\sin\tilde{\varphi},\qquad y^3=\cos\tilde{\theta}\, .
\end{equation}
The complex coordinate $z$ on $\Sigma$ is defined by $z=x^1+ix^2$. 
\\
\indent The four scalars from $SO(4,1)/SO(4)$ can be parametrized by $m_\Lambda$, $\Lambda=0,1,2,3,4$, satisfying
\begin{equation}
m_\Lambda \eta^{\Lambda\Sigma}m_\Sigma=-1 \qquad \textrm{and}\qquad m^\Lambda m^\Sigma=\delta^{\alpha\beta}{L^\Lambda}_\alpha {L^\Sigma}_\beta-\eta^{\Lambda\Sigma}
\end{equation}
with ${L^\Lambda}_{\ul{\Lambda}}$ being the coset representative used in the main text. We have renamed various indices used in \cite{Henning_Malek_AdS7_6} in order to match with the $F(4)$ gauged supergravity in \cite{F4SUGRA1} and \cite{F4SUGRA2}. In particular, it is useful to note the relation
\begin{equation}
X=e^\sigma,\qquad B_{(2)}=-B,\qquad R=\frac{\sqrt{2}}{m}=\frac{3\sqrt{2}}{g_1}
\end{equation}  
with $R$ being the $\tilde{S}^2$ radius. Recall that $SO(2)_R$ invariant coset representative is given by
\begin{equation}
L=e^{\phi_0Y_{01}+\phi_2Y_{04}},
\end{equation}
we find 
\begin{equation}
m_\Lambda=(\sinh\phi_0,0,0,\cosh\phi_0\cosh\phi_2,\cosh\phi_0\sinh\phi_2).
\end{equation}
\indent With all these, we can write down the ten-dimensional metric and other fields in type IIB theory as follows
\begin{eqnarray}
d\hat{s}^2&=&\frac{8}{g_1}c_6(3\rho)^{\frac{5}{4}}\left(\frac{|dk|^2}{\bar{\Delta}}\right)^{\frac{3}{4}}\left[\frac{g_1^2\bar{\Delta}}{6\rho|dk|^2}ds^2+\frac{\widetilde{\Delta}e^{-2\sigma}}{\rho^2|dk|^2}dk^adp_a \right.\nonumber \\
& &+e^{2\sigma}\left(d\tilde{\theta}^2+\sin^2\tilde{\theta}d\tilde{\varphi}^2+w^2-\frac{1}{\rho^2}p_a p_b n^a n^b\right)\nonumber \\
& &\left. -\frac{3}{\rho}e^{-2\sigma}n_a(\sinh\phi_0dk^a-\cosh\phi_0\cosh\phi_2\cos\tilde{\theta}dp^a)\right],\\
H^{ab}&=&\frac{e^{4\sigma}p^ap^b|dk|}{\sqrt{3\rho\bar{\Delta}}}+\sqrt{\frac{3\rho}{\bar{\Delta}}}\left[\cosh\phi_0\sinh\phi_2\pd_ck^a\pd^cp^b\right. \nonumber \\
& &\left.+n^{ac}(\cosh\phi_0\cosh\phi_2\cos\tilde{\theta}\pd_cp^b-\sinh\phi_0\pd_c k^b)\right],\\
C^a_{(2)}&=&-\frac{24c_6}{g_1^2}\textrm{vol}_{\tilde{S}^2}(k^a+L^a)\nonumber \\
& &-\frac{24c_6}{g_1^2}\frac{|dk|^2}{\bar{\Delta}}\cosh\phi_0\cosh\phi_2\sin^2\tilde{\theta}\sigma^a\wedge d\tilde{\varphi}-4c_6p^aB,\nonumber \\
& &\\
F_{(5)}&=&F_{(2,3)}+F_{(3,2)}+F_{(4,1)}
\end{eqnarray}
with
\begin{eqnarray}
F_{(2,3)}&=&\frac{288c_6^2}{g_1^3}|dk|\left[\frac{1}{3}\frac{g_1\rho |dk|}{\bar{\Delta}}e^{4\sigma}p_aB\wedge \rho^a\wedge \textrm{vol}_{\tilde{S}^2}-\frac{|dk|}{\bar{\Delta}}\cosh\phi_0\cosh\phi_2\sin^2\tilde{\theta}\times  \right.\nonumber \\
& &\left.\left(2g_1\rho|dk|\sinh\phi_0+\frac{1}{3}g_1e^{4\sigma}p_ap_bn^{ac}\pd_ck^b\right)B\wedge\textrm{vol}_\Sigma\wedge d\tilde{\varphi} \right],\nonumber
\end{eqnarray}
\begin{eqnarray} 
F_{(3,2)}&=&\frac{288c_6^2}{g_1^2}\frac{\rho^2|dk|^2}{\bar{\Delta}}dB\wedge \left[\cosh\phi_0\cosh\phi_2\sin^2\tilde{\theta}w\wedge d\tilde{\varphi} \right.\nonumber \\
& &\left.-(\cosh^2\phi_0\sinh^2\phi_2-\sinh^2\phi_0-\cosh^2\phi_0\cosh^2\phi_2\cos^2\tilde{\theta})\textrm{vol}_{\tilde{S}^2} \right]\nonumber \\
& &-\frac{288c_6^2|dk|}{g_1^2}e^{-4\sigma}*_6dB\wedge \textrm{vol}_\Sigma\nonumber \\
&=&*_{10}F_{(3,2)},\nonumber \\
F_{(4,1)}&=&*_{10}F_{(2,3)}\, .
\end{eqnarray}
Various quantities appearing in the above equations are defined by
\begin{eqnarray}
\bar{\Delta}&=&e^{4\sigma}|dk|p_ap_b\left[\cosh\phi_0\sinh\phi_2\pd_ck^a\pd^cp^b\right.\nonumber \\
& &\left.+n^{ac}(\cosh\phi_0\cosh\phi_2\cos\tilde{\theta}\pd_cp^b-\sinh\phi_0\pd_ck^b)\right]\nonumber \\
& &+3\rho|dk|^2(\cosh^2\phi_0\sinh^2\phi_2-\sinh^2\phi_0-\cosh^2\phi_0\cosh^2\phi_2\cos^2\tilde{\theta}),\nonumber \\
\widetilde{\Delta}&=&3\rho\cosh\phi_0\sinh\phi_2|dk|^2+e^{4\sigma}|dk|p_ap_b\pd_ck^a\pd^cp^b,\nonumber \\
w&=&\frac{1}{3\rho^2}p_a\sigma^a-\cosh\phi_0\cosh\phi_2\sin\tilde{\theta}d\tilde{\theta},
\nonumber \\
\sigma^a&=&3\rho(\cosh\phi_0\sinh\phi_2n^a-\sinh\phi_0dp^a-\cosh\phi_0\cosh\phi_2\cos\tilde{\theta}dk^a)\nonumber \\
& &-e^{4\sigma}p^ap_b*_2n^b,\nonumber \\
|dk|&=&\frac{1}{2}\pd_ak_b\pd^ak^b,\qquad n^a=\frac{1}{2}\mathfrak{g}\pd f^ad\bar{z}+\frac{1}{2}\bar{\mathfrak{g}}\bar{\pd} \bar{f}^adz,\nonumber \\
L^a&=&\frac{e^{4\sigma}\rho |dk|p_b}{\bar{\Delta}}\left[\cosh\phi_0\sinh\phi_2\pd_ck^b\pd^cp^a\right. \nonumber \\
& &\left.+n^{bc}(\cosh\phi_0\cosh\phi_2\cos\tilde{\theta}\pd_ck^a-\sinh\phi_0\pd_ck^a)\right],\nonumber \\
\rho^a&=&\cosh\phi_0\sinh\phi_2dk^a-\cosh\phi_0\cosh\phi_2\cos\tilde{\theta}n^a-\sinh\phi_0*_2n^a
\end{eqnarray}
with $*_2$ being the Hodge duality on the flat $\Sigma$. $c_6$ is a constant, and $\mathfrak{g}$ is a $U(1)$ phase.  
\\
\indent The ten-dimensional dilaton $\Phi$ and axion $C_0$ are encoded in the matrix $H_{ab}$ as
\begin{equation}
H_{ab}=\frac{1}{\textrm{Im}\, \tau}\left(\begin{array}{cc}
|\tau|^2 & \textrm{Re}\, \tau \\
\textrm{Re}\, \tau & 1
\end{array}\right)\label{Hab}
\end{equation}
with $\tau=e^\Phi+iC_0$.
%%%%%%%%%%%%%%%%%%%%%%%%%%%%%%%%%%%%%%%%%%%%%%%%%%%%%%%%%%%%%%%%%%%%%%%%%%%%%%%%%%%%%%%%%%%%%%%%%%%%%%%%%%%%%%%%%%%%%%%%%%%%%%%%%%%%%%%%%


\begin{thebibliography}{99}
\bibitem{maldacena} J. M. Maldacena, ``The large $N$ limit of
superconformal field theories and supergravity'', Adv. Theor. Math.
Phys. \textbf{2} (1998) 231-252, arXiv: hep-th/9711200.
\bibitem{Gubser_AdS_CFT} S. S. Gubser, I. R. Klebanov and A. M. Polyakov, ``'', Phys. Lett. \textbf{B428} (1998) 105-114, arXiv: hep-th/9802.109.
\bibitem{Witten_AdS_CFT} E. Witten, ``Anti De Sitter Space and holography'', Adv. Theor. Math. Phys. \textbf{2} (1998) 253-291, arXiv: 9802150.
\bibitem{defect1} O. DeWolfe, D. Z. Freedman, and H. Ooguri, “Holography and defect conformal field theories,” Phys.
Rev. D66 (2002) 025009, arXiv:hep-th/0111135 [hep-th].
\bibitem{defect2} C. Bachas, J. de Boer, R. Dijkgraaf, and H. Ooguri, “Permeable conformal walls and holography,”
JHEP 06 (2002) 027, arXiv:hep-th/0111210 [hep-th].
\bibitem{defect3} O. Aharony, O. DeWolfe, D. Z. Freedman, and A. Karch, “Defect conformal field theory and locally
localized gravity,” JHEP 07 (2003) 030, arXiv:hep-th/0303249 [hep-th].
\bibitem{boundary_defect} T. Quella and V. Schomerus, ``Symmetry Breaking Boundary States and Defect Lines'', JHEP 06 (2002) \textbf{028}, arXiv: hep-th/0203161.
\bibitem{N08_defect} J. A. Harvey and A. B. Royston, ``Gauge/gravity duality with a chiral N=(0,8) string defect'', JHEP 08 (2008) \textbf{006}, arXiv: 0804.2854.
\bibitem{quantum_AdS_dCFT1} I. Buhl-Mortensen, M. de Leeuw, A. C. Ipsen, C. Kristjansen and M. Wilhelm, ``A Quantum CHeck of AdS/dCFT'', JHEP 01 (2017) \textbf{098}, arXiv: 1611.04603.
\bibitem{quantum_AdS_dCFT2} A. G. Grau, C. Kristjansen and M. Wilhelm, ``A Quantum CHeck of Non-Supersymmetric AdS/dCFT'', JHEP 01 (2019) \textbf{007}, arXiv: 1810.11463.
\bibitem{surface_AdS3} F. Faedo, Y. Lozano and N. Petri, ``Searching for surface defect CFTs within $AdS_3$'', JHEP 11 (2020) \textbf{052}, arXiv: 2007.16167.
\bibitem{AdS2_4D_defect} Y. Lozano, N. Petri and C. Risco, ``New $AdS_2$ supergravity duals of 4D SCFTs with defects'', JHEP 10 (2021) \textbf{217}, arXiv: 2107.12277.
\bibitem{AdS2_defect} Y. Lozano, N. Petri and C. Risco, ``$AdS_2$ near-horizons, defects and string dualities'', Phys. Rev. \textbf{D107} (2023) 106012, arXiv: 2212.11095. 
\bibitem{correlator_defect} G. Georgiou, G. Linardopoulos and D. Zoakos, ``Holographic correlators of semiclassical states in defect CFTs'', Phys. Rev. \textbf{D108} (2023) 046016, arXiv: 2304.10434.
\bibitem{F4SUGRA1} R. D' Auria, S. Ferrara and S. Vaula, ``Matter coupled $F(4)$ supergravity and the AdS$_6$/CFT$_5$
correspondence'', JHEP 10 (2000) \textbf{013}, arXiv:
hep-th/0006107.
\bibitem{F4SUGRA2} L. Andrianopoli, R. D' Auria and S. Vaula, ``Matter coupled $F(4)$ gauged supergravity
Lagrangian'', JHEP 05 (2001) \textbf{065}, arXiv: hep-th/0104155.
\bibitem{F4_Romans} L. J. Romans, ``The $F(4)$ gauged supergravity in
six-dimensions'', Nucl. Phys \textbf{B269} (1986) 691.
\bibitem{AdS2_F4_Dibitetto} G. Dibitetto and N. Petri, ``$AdS_2$ solutions and their massive IIA origin'', JHEP 05 (2019) \textbf{107}, arXiv: 1811.11572.
\bibitem{surface_defect_F4_Dibitetto} G. Dibitetto and N. Petri, ``Surface defects in the $D4-D8$ brane system'', JHEP 01 (2019) \textbf{193}, arXiv: 1807.07768.
\bibitem{line_defect_F4_Gutperle} K. Chen and M. Gutperle, ``Holographic line defects in $F(4)$ gauged supergravity'', Phys. Rev. \textbf{D100} (2019) 126015, arXiv: 1909.11127.
\bibitem{F4_form_mIIA} M. Cvetic, H. Lu and C. N. Pope, ``Gauged six-dimensional supergravity from massive type IIA'', Phys. Rev. Lett. \textbf{83} (1999) 5226, arXiv: hep-th/9906221.
\bibitem{Christoph1} C. F. Uhlemann, ``Wilson loops in 5d long quiver gauge theories'', JHEP 09 (2020) \textbf{145}, arXiv: 2006.01142.
\bibitem{Christoph2} M. Gutperle and C. F. Uhlemann, ``Surface defects in holographic 5d SCFTs'', JHEP 04 (2021) \textbf{134}, arXiv: 2012.14547.
\bibitem{Christoph3} L. Santilli and C. F. Uhlemann, ``3d defects in 5d: RG flows and defect F-maximization'', JHEP 06 (2023) \textbf{136}, arXiv: 2305.01004.
\bibitem{F4_flow} P. Karndumri, ``Holographic RG flows in six dimensional F(4) gauged
supergravity'', JHEP 01 (2013) \textbf{134}, Erratum-ibid. JHEP 06 (2015) \textbf{165}, arXiv: 1210.8064.
\bibitem{5DSYM_from_F4} P. Karndumri, ``Gravity duals of 5D N=2 SYM from F(4) gauged supergravity'',
Phys. Rev. \textbf{D90} (2014) 086009, arXiv: 1403.1150.
\bibitem{6D_twist} P. Karndumri, ``Twisted compactification of $N = 2$ 5D SCFTs to three and two dimensions from $F(4)$ gauged supergravity'', JHEP 09 (2015) \textbf{034}, arXiv: 1507.01515.
\bibitem{AdS6_BH} P. Karndumri, ``New supersymmetric $AdS_6$ black holes from matter-coupled $F(4)$ gauged supergravity'', arXiv: 2403.01746.
\bibitem{6D_Janus_RG} P. Karndumri, ``Janus and RG-flow interfaces from matter-coupled $F(4)$ gauged supergravity'', arXiv: 2405.17169.
\bibitem{6D_Janus} M. Gutperle, J. Kaidi and H. Raj, ``Janus solutions in six-dimensional gauged supergravity'', JHEP 12 (2017) \textbf{018}, arXiv: 1709.09204.
\bibitem{AdS6_BH_Zaffaroni} S. M. Hosseini, K. Hristov, A. Passias, A. Zaffaroni,  ``6D attractors and black hole microstates'', JHEP 12 (2018) \textbf{001}, arXiv: 1809.10685.
\bibitem{AdS6_BH_Minwoo} M. Suh, ``Supersymmetric $AdS_6$ black holes from matter coupled $F(4)$ gauged supergravity'', JHEP 02 (2019) \textbf{108}, arXiv: 1810.00675.
\bibitem{7D_sol_Dibitetto} G. Dibitetto and N. Petri, ``BPS objects in $D=7$ supergravity and their M-theory
origin'', JHEP 12 (2017) \textbf{041}, arXiv: 1707.06152.
\bibitem{7D_N2_DW_3_form} P. Karndumri and P. Nuchino, ``Supersymmetric solutions from matter-coupled $7D$ $N=2$ gauged supergravity'', Phys. Rev. \textbf{D98} (2018) 086012, arXiv: 1806.04064.
\bibitem{6D_surface_Dibitetto} G. Dibitetto and N. Petri, ``6d surface defects from massive type IIA'', JHEP 01 (2018) \textbf{039}, arXiv: 1707.06154.
\bibitem{7D_N4_DW_3_form} P. Karndumri and P. Nuchino, ``Supersymmetric solutions of 7D maximal gauged supergravity'', Phys. Rev. \textbf{D101} (2020) 086012, arXiv: 1910.02909.
\bibitem{7D_defect_Nicolo} Y. Lozano, N. T. Macpherson, N. Petri and C. Risco, ``New AdS$_3$/CFT$_2$ pairs in massive IIA with $(0,4)$ and $(4,4)$ supersymmetries'', JHEP 09 (2022) \textbf{130}, arXiv: 2206.13541.
\bibitem{Henning_Malek_AdS7_6} E. Malek, H. Samtleben and V. V. Camell, ``Supersymmetric $AdS_7$ and $AdS_6$ vacua and their consistent truncations with vector multiplets'', JHEP 04 (2019) \textbf{088}, arXiv: 1901.11039.
\bibitem{Gubser_singularity} S. S. Gubser, ``Curvature singularities: the good, the bad and the naked'', Adv. Theor.
Math. Phys. \textbf{4} (2000) 679-745.
\bibitem{maldacena_nogo} J. Maldacena and C. Nunez, ``Supergravity description of field theories on
curved manifolds and a no go theorem'', Int. J. Mod. Phys. \textbf{A16} (2001) 822,
arXiv: hep-th/0007018.
\bibitem{IIB_AdS6} E. D'Hoker, M. Gutperle and C. F. Uhlemann, ``Holographic duals for five-dimensional superconformal quantum field theories'', Phys. Rev. Lett. 118 (2017) 101601, arXiv: 1611.09411.
\bibitem{04_defect_Nicolo} F. Faedo, Y. Lozano and N. Petri, ``New $N=(0,4)$ AdS$_3$ near-horizons in Type IIB'', JHEP 04 (2021) \textbf{028}, arXiv: 2012.07148.
\end{thebibliography}
\end{document}